\documentclass[aps,prl,amsmath,twocolumn,preprintnumber,superscriptaddress,showpacs,notitlepage]{revtex4-2}
\usepackage[T1]{fontenc}
\usepackage[latin9]{inputenc}
\DeclareMathAlphabet\mathbfcal{OMS}{cmsy}{b}{n}

\makeatletter

\usepackage{mathptmx,newtxtext,newtxmath,xspace}
\usepackage{amsbsy,bm,bbold}
\usepackage{graphicx,color,xcolor,epsfig,rotate}
\usepackage{fancyhdr}
\usepackage[colorlinks=true, 
            linkcolor=blue, 
            urlcolor=blue,
            citecolor=blue]{hyperref}
\usepackage{soul}
\usepackage{cancel}
\makeatother

\usepackage{babel}
\begin{document}
\title{Nonperturbative Semiclassical Spin Dynamics for Ordered Quantum Magnets}
\author{Hao~Zhang}
\affiliation{Department of Physics and Astronomy, The University of Tennessee,
Knoxville, Tennessee 37996, USA}
\author{Tianyue~Huang}
\affiliation{Institute of Physics, \'{E}cole Polytechnique F\'{e}d\'{e}rale de Lausanne (EPFL), CH-1015 Lausanne, Switzerland}
\author{Allen~O.~Scheie}
\affiliation{MPA-Q, Los Alamos National Laboratory, Los Alamos, New Mexico 87545, USA}
\author{Mengze~Zhu}
\affiliation{Laboratory for Solid State Physics, ETH Z\"{u}rich, 8093 Z\"{u}rich, Switzerland}
\author{Tao~Xie}
\affiliation{Center for Neutron Science and Technology, Guangdong Provincial Key Laboratory of Magnetoelectric Physics and Devices,
School of Physics, Sun Yat-sen University, Guangzhou, Guangdong 510275, China}
\author{N.~Murai}
\affiliation{J-PARC Center, Japan Atomic Energy Agency, Tokai, Ibaraki 319-1195, Japan}
\author{S.~Ohira-Kawamura}
\affiliation{J-PARC Center, Japan Atomic Energy Agency, Tokai, Ibaraki 319-1195, Japan}
\author{Andrey~Zheludev}
\affiliation{Laboratory for Solid State Physics, ETH Z\"{u}rich, 8093 Z\"{u}rich, Switzerland}
\author{Andreas~M.~L\"{a}uchli}
\affiliation{Laboratory for Theoretical and Computational Physics, Paul Scherrer Institut, CH-5232 Villigen-PSI, Switzerland}
\affiliation{Institute of Physics, \'{E}cole Polytechnique F\'{e}d\'{e}rale de Lausanne (EPFL), CH-1015 Lausanne, Switzerland}
\author{Cristian~D.~Batista}
\affiliation{Department of Physics and Astronomy, The University of Tennessee,
Knoxville, Tennessee 37996, USA}
\affiliation{Quantum Condensed Matter Division and Shull-Wollan Center, Oak Ridge
National Laboratory, Oak Ridge, Tennessee 37831, USA}

\date{\today}
\begin{abstract}
In ordered quantum magnets where interactions between elementary excitations dominate over their kinetic energy, perturbative approaches often fail, making non-perturbative methods essential to capture spectral features such as bound states and the redistribution of weight within excitation continua. Although an increasing number of experiments report anomalous spin excitation continua in such systems, their microscopic interpretation remains an open challenge. Here, we investigate the spin dynamics of the triangular-lattice antiferromagnet in its 1/3-plateau phase using two complementary non-perturbative approaches: exact diagonalization in a truncated Hilbert space for a gas of elementary excitations (THED) and matrix product state (MPS) simulations. Alongside cross-validation between these methods, we benchmark our results against inelastic neutron scattering (INS) data. The THED analysis confirms the presence of two-magnon bound states and identifies the anomalous scattering continuum observed in both MPS and INS as a two-magnon resonance, arising from hybridization between the bound state and the two-magnon continuum. Furthermore, THED reveals bound states overlapping with the continuum, enriching the interpretation of continuum anomalies. More broadly, THED provides a robust framework for investigating anomalous spin excitation continua and bound-state effects in other materials with gapped spectra. Its combination of accuracy and computational efficiency makes it a powerful tool for extracting reliable microscopic models in semiclassical regimes.
\end{abstract}
\maketitle

Modern condensed matter physics rests on the notion of quasiparticles, which allows for the recasting of complex many-body dynamics in terms of weakly interacting collective excitations. In ordered quantum magnets, these excitations manifest as quantized spin waves, or magnons. When magnon-magnon interactions are weak, diagrammatic perturbation theory (loop expansion) provides a powerful framework for describing their dynamics, accurately capturing features such as spectral renormalization and intrinsic linewidth broadening~\cite{ZhitomirskyME2013,ChernyshevAL2009,PlumbKW2016,KamiyaY2018a,DoSH2021a}. 

As interactions become comparable to or exceed the magnon kinetic energy, the system can no longer be treated as a weakly interacting gas and perturbative approaches break down. In this regime, quasiparticles can form bound states or resonances, undergo strong renormalization, or decay via channels involving bound states~\cite{VerresenR2019,DallyRL2020,BaiX2021,BaiX2023,LegrosA2021,ShengJ2025a}. These strong interactions substantially reshape the structure of the multi-magnon continuum, making it essential to accurately capture non-perturbative effects in order to extract reliable microscopic models. Furthermore, the softening of bound states under variation of Hamiltonian parameters often signals the onset of exotic multipolar orderings~\cite{Wang2017,ShengJ2025a,HuangQ2025a}.  
Despite a growing number of experiments reporting anomalous spin excitation continua in ordered magnets~\cite{KamiyaY2018a,XieT2023,ScheieAO2024,ShengJ2025b}, their microscopic origin remains largely unresolved, partly due to the scarcity of non-perturbative theoretical tools capable of capturing these complex many-body dynamics.

Recent methodological advances in the time evolution of matrix product states (MPS)~\cite{haegeman_time-dependent_2011,haegeman_unifying_2016,vanderstraeten_tangent-space_2019,paeckel_time-evolution_2019} have enabled the calculation of spectral functions that can be directly compared with experimental measurements, such as inelastic neutron scattering (INS). These techniques have proven highly effective in capturing complex non-perturbative effects in ordered quantum magnets~\cite{XieT2023,Hernandez2025,ShengJ2025b}. However, being purely numerical approaches, MPS-based methods often provide limited interpretative insight into the underlying physical mechanisms. Moreover, accurate MPS simulations in two or more spatial dimensions remain computationally challenging, limiting the applicability of these methods for extracting effective microscopic models of real materials.

In this article, we test and extend the truncated Hilbert space exact diagonalization (THED) method, originally developed to capture non-perturbative single-particle effects in the effective spin-one magnet $\mathrm{FeI_2}$~\cite{LegrosA2021,BaiX2023}, to investigate multi-particle continua in more general gapped ordered quantum magnets. By restricting the full Hilbert space to sectors containing a limited number of magnons, THED captures few-body non-perturbative effects while reducing the computational complexity from exponential to polynomial scaling. This approach enables the resolution of non-perturbative spectral features with clear physical interpretation, provided the system possesses a finite single-magnon gap. Our calculations are carried out using a recently developed general-purpose implementation of the method~\cite{ZhangH2025}.

\begin{figure*}
    \centering
    \includegraphics[width=1.8\columnwidth]{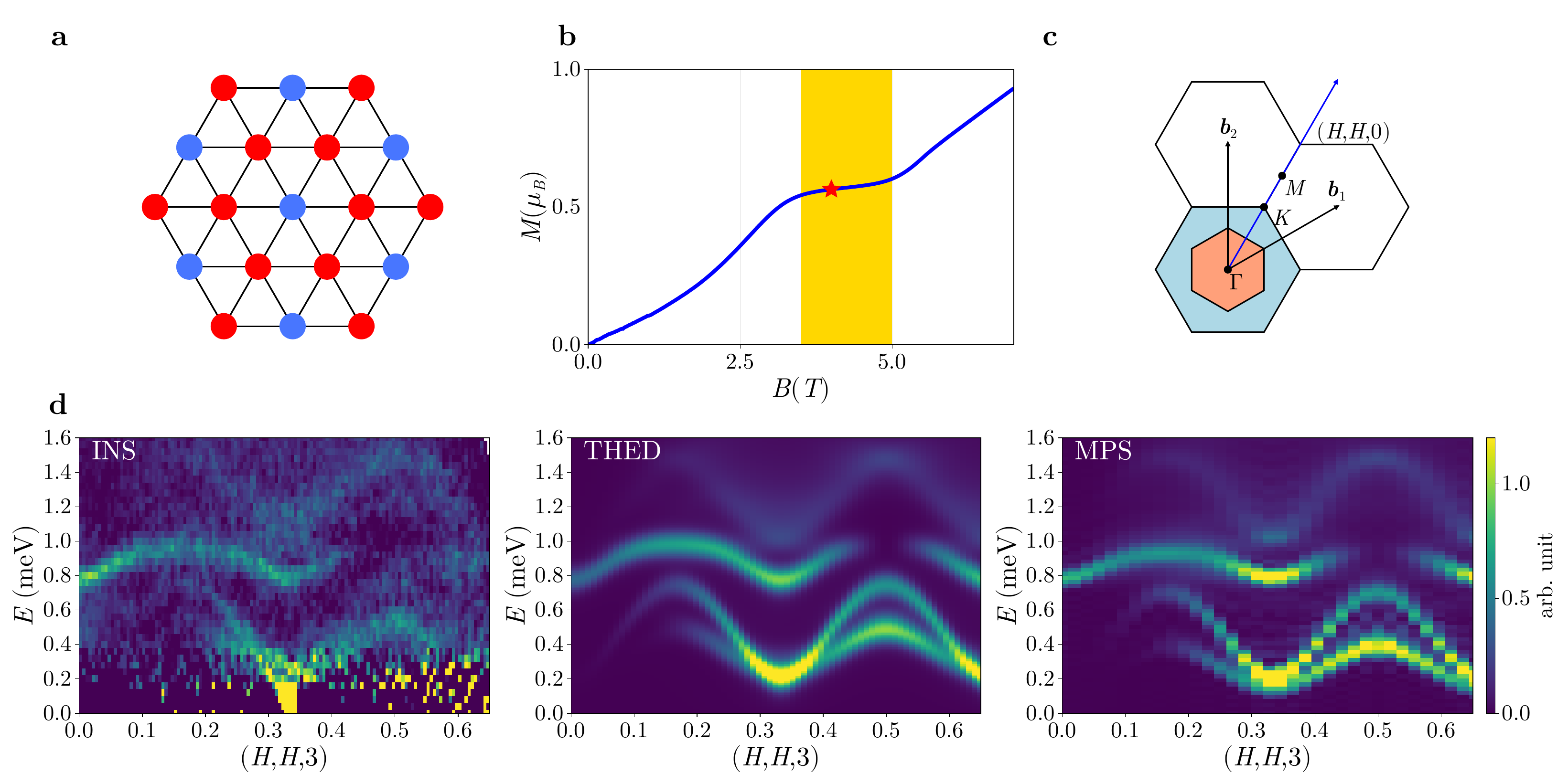}
    \caption{{\bf The 1/3 plateau phase of the TLAFM}. {\bf a} The up-up-down (UUD) magnetic order represented by red for spin-up and blue for spin-down states. {\bf b} The magnetization curve of $\mathrm{KYbSe_2}$ under an external magnetic field, highlighting the 1/3 plateau. {\bf c} The Brillouin zone of the triangular lattice (light blue shading) and the reduced Brillouin zone corresponding to the UUD order (light red shading), along with the chosen path. {\bf d} Comparison of neutron scattering intensities for $\mathrm{KYbSe_2}$ at $B=4\ \mathrm{T}$ (marked by a star in panel b) with simulated results from the truncated Hilbert space exact diagonalization (THED) method and the time evolution of the matrix product state (MPS).}
    \label{fig:1}
\end{figure*}

Here, we test our method on the spin-1/2 triangular-lattice Heisenberg antiferromagnet (TLHAFM) in the presence of a magnetic field, where a robust 1/3-magnetization plateau emerges. This field-induced phase, originally predicted by Chubukov and Golosov~\cite{ChubukovAV1991}, hosts a distinctive ``up-up-down'' (UUD) spin configuration~(Fig.~\ref{fig:1}{\bf a}) and has since been observed across a wide range of spin-1/2 triangular-lattice compounds~\cite{OnoT2003,InamiT1996,ShirataY2012,ZhouHD2012,SusukiT2013,KoutroulakisG2015,QuirionG2015,MaJ2016,SeraA2016,ItoS2017a}. Long recognized as a hallmark of frustrated magnetism, the plateau not only reflects the stabilization of quantum order~\cite{AliceaJ2009,ColettaT2013,Starykh2015,ColettaT2016,Syromyatnikov2023} but also provides a fertile ground for exploring strongly interacting spin dynamics. A persistent puzzle in this context is the presence of an extended excitation continuum above the single-magnon modes that evades explanation by conventional spin-wave theory~\cite{KamiyaY2018a,ScheieAO2024}. 

While recent MPS simulations have succeeded in reproducing these broad spectral features~\cite{XieT2023}, they offer limited insight into the underlying microscopic processes and the nature of the modes. The origin of the continuum therefore remains an open question, pointing to the need for new theoretical frameworks capable of capturing the inherently non-perturbative nature of spin excitations in this regime.

Beyond its intrinsic interest, the study of excitation dynamics in the 1/3-magnetization plateau phase is also directly relevant to a family of spin-1/2 $J_1$-$J_2$ triangular-lattice compounds, $A\mathrm{YbSe_2}$ (with $A=\mathrm{Na,K,Cs}$)~\cite{Dai2021,ScheieAO2024d,ScheieAO2024b,XieT2023}, which reside near a quantum critical regime separating magnetically ordered states from a putative spin-liquid ground state~\cite{ScheieAO2024d}. While semiclassical methods are often appropriate in the field-induced UUD phase, capturing the essential features of the excitation spectrum requires incorporating non-perturbative effects. Accurate modeling in this regime is crucial not only for understanding the anomalous spin dynamics but also for reliably extracting microscopic Hamiltonian parameters, which serve as inputs to zero-field studies probing quantum criticality~\cite{ScheieAO2024b}.

In this work, we perform parallel simulations using both the THED and MPS methods to enable cross-validation. In the THED approach, we include contributions up to the two-magnon sector to capture the anomalous continua observed in INS experiments, accounting for the finite single-magnon gap generated by quantum fluctuations in the UUD phase~\cite{AliceaJ2009,KamiyaY2018a}. For the MPS simulations, we carry out calculations in a cylindrical geometry with moderate bond dimensions and cluster sizes, taking advantage of the weak entanglement characteristic of the UUD phase. Technical details of both approaches are provided in the ``Methods'' section.

By combining these two approaches and benchmarking them against INS data, we identify the anomalous continuum in the UUD phase of TLHAFM as a two-magnon resonance arising from the hybridization between a bound state and the multi-magnon continuum. Furthermore, the interpretability afforded by the THED method allows us to uncover bound states embedded within the continuum, which are reminiscent of the bound states in the continuum (BICs) of photonic systems~\cite{MarinicaDC2008,KangM2023}.

Beyond the INS examples presented in this work, the predictions of the THED method can be directly compared with a range of other spectroscopic probes of spin dynamics for ordered quantum magnets, including Raman scattering, electron spin resonance, optical conductivity, and resonant inelastic X-ray scattering.

More broadly, the THED framework--performing exact diagonalization while constraining the number of elementary excitations--can be applied to any $N$-level system exhibiting semiclassical behavior. This includes atomic gases in optical lattices~\cite{Bloch2008} serving as quantum simulators, quantum optical systems~\cite{HsuCW2016}, and orbital degrees of freedom in solids~\cite{Khaliullin2005,Liu2025}. In these contexts, THED offers a computationally efficient approach to capturing non-perturbative phenomena while keeping the numerical complexity tractable.

\section*{Results}
The spin Hamiltonian for the triangular lattice antiferromagnet (TLAFM) is given by
\begin{equation}
    \mathcal{H} = \sum_{\langle i,j \rangle} J_{ij} (S_i^x S_j^x + S_i^y S_j^y + \Delta_{ij} S_i^z S_j^z) - \mu_{B} \bm{B}^{T}  \bm{g}  \sum_j \bm{S}_j.
    \label{eq:model}
\end{equation}
The first term describes the antiferromagnetic (AFM) exchange interaction, with $J_{ij}>0$ acting on the first- and second-nearest neighbor bonds, while $\Delta_{ij}$ is the dimensionless exchange anisotropy. The second term represents the Zeeman interaction, expressed as the contraction of the external magnetic field $\bm{B}$ with the $g$-tensor $\bm{g}$ and the spin operator $\bm{S}_j$ of site $j$.  Throughout the main text, we either assume that the external field is aligned with the easy-axis (taken to be the $z$-axis), or we employ an isotropic Heisenberg model with $\Delta_{ij} = 1$, which preserves the U(1) symmetry of the Hamiltonian for arbitrary field directions. Our main goal is to investigate the low-energy spin excitations of this model in the field-induced three-sublattice UUD ordered phase. 

The compound $\mathrm{KYbSe_2}$ (KYS) provides a representative realization of the model $\mathcal{H}$ with parameters $J_1 = 0.456 \ \text{meV}$, $J_2/J_1 \simeq 0.043$, and $\Delta_{ij} = 1$. Figure~\ref{fig:1}{\bf b} shows the magnetization curve of KYS under an external magnetic field applied along the $x$-axis, with the 1/3-plateau phase highlighted in orange. Our main results for KYS in this plateau phase at $B = 4~\mathrm{T}$ are presented in Fig.~\ref{fig:1}{\bf d}, where we compare the INS cross-section with simulations performed using THED and MPS methods along the $(H,H,3)$ direction (see Fig.~\ref{fig:1}{\bf c} for the momentum path).  Notably, both methods reproduce the key spectral features observed in the INS data.

Three bright branches are observed below 1.0~meV, corresponding to collective single-magnon excitations originating from the three sublattices. These modes are well described by including the leading $1/S$ correction to linear spin wave theory (LSWT)~\cite{AliceaJ2009,Starykh2015,KamiyaY2018a}. However, as we will show below, the spectral weight between 1.0 and 1.6~meV exhibits a structured two-magnon continuum that cannot be captured by any finite-order $1/S$ expansion~\cite{KamiyaY2018a,ScheieAO2024}. Remarkably, both the THED and MPS methods accurately reproduce these non-perturbative features. A central message of this work is that, beyond capturing such effects, the THED approach provides complementary insights to MPS by elucidating the internal structure of the two-magnon continuum.

\begin{figure}
    \centering
    \includegraphics[width=1\columnwidth]{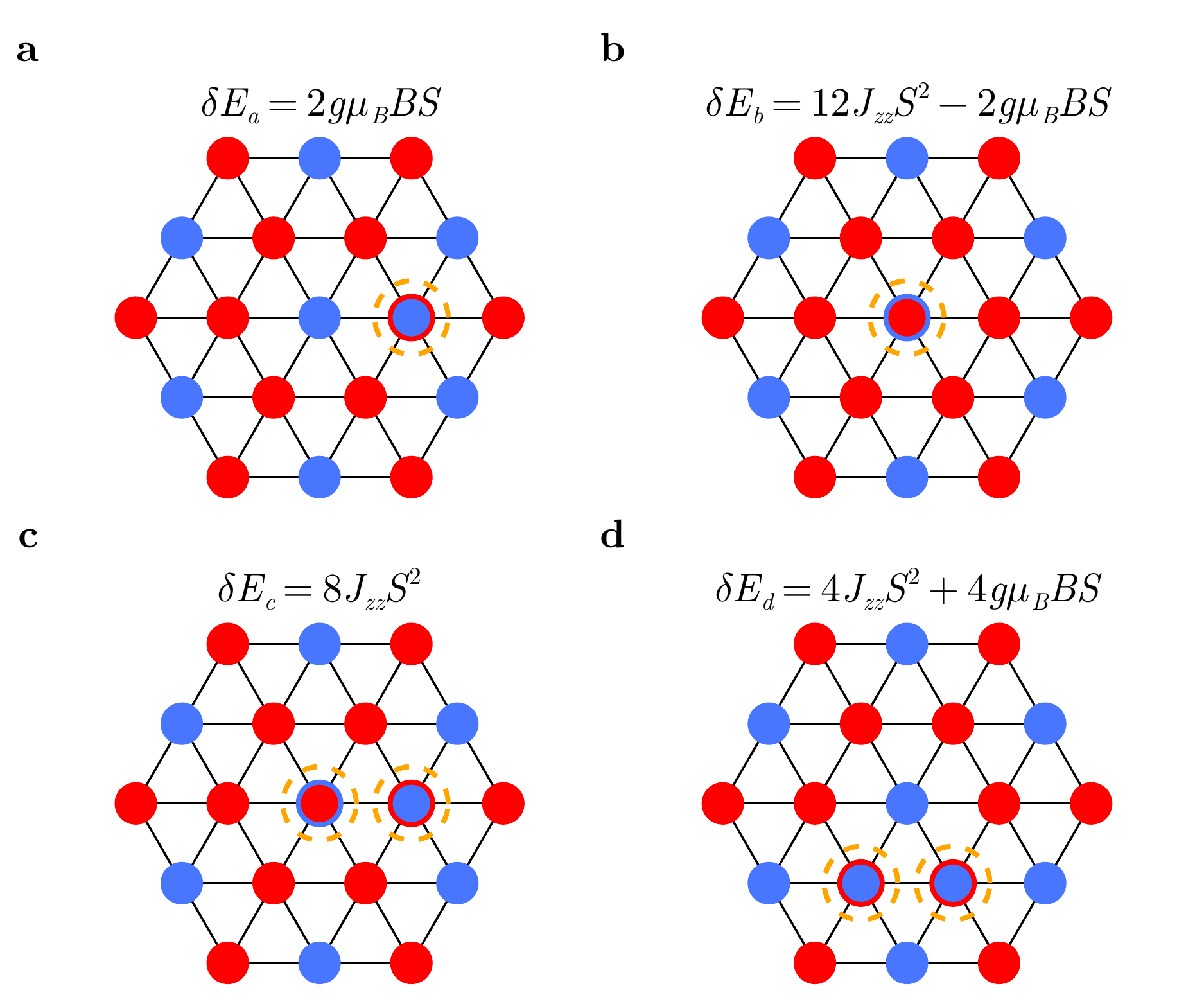}
    \caption{{\bf Excitations of the classical Ising model up to two spin flips.} {\bf a}--{\bf d} Red (blue) circles represent spin-up (spin-down) states, while the orange dashed circle highlights the position of a spin flip. The energy cost associated with each excitation is shown above each panel.}
    \label{fig:2}
\end{figure}

To understand the nature of the two-magnon excitations, we begin by considering the classical Ising limit ($\Delta_{ij} \to \infty$). For simplicity, we restrict the discussion to nearest-neighbor interactions with $J_{zz} = J_1$, unless otherwise noted. The inclusion of a small $J_2/J_1$ ratio leads to only quantitative modifications, without altering the qualitative physics. Figure~\ref{fig:2} summarizes the relevant excitations involving up to two spin flips in the classical Ising limit.

We first analyze two-spin-flip processes occurring on both the ``up'' and ``down'' sublattices. As illustrated in Fig.~\ref{fig:2}, flipping two neighboring spins incurs a lower energy cost than flipping them farther apart, specifically $\delta E_c < \delta E_a + \delta E_b$, leading to the formation of two-magnon bound states. These excitations carry a total spin change $\Delta S^z = 0$, making them observable via INS. As we show below, the anomalous continua observed in the INS spectra of materials near the isotropic limit can be traced back to these bound-state excitations in the Ising limit.

Another two-spin-flip process, shown in Fig.~\ref{fig:2}{\bf d}, corresponds to a distinct bound state with $\Delta S^z = -2$. This quadrupolar excitation is invisible to INS within the current model. However, as demonstrated in  Supplementary Sec.~IV, such bound states can acquire finite spectral weight if the U(1) symmetry along the easy-axis in Eq.~\eqref{eq:model} is broken.

\begin{figure*}
    \centering
    \includegraphics[width=1.5\columnwidth]{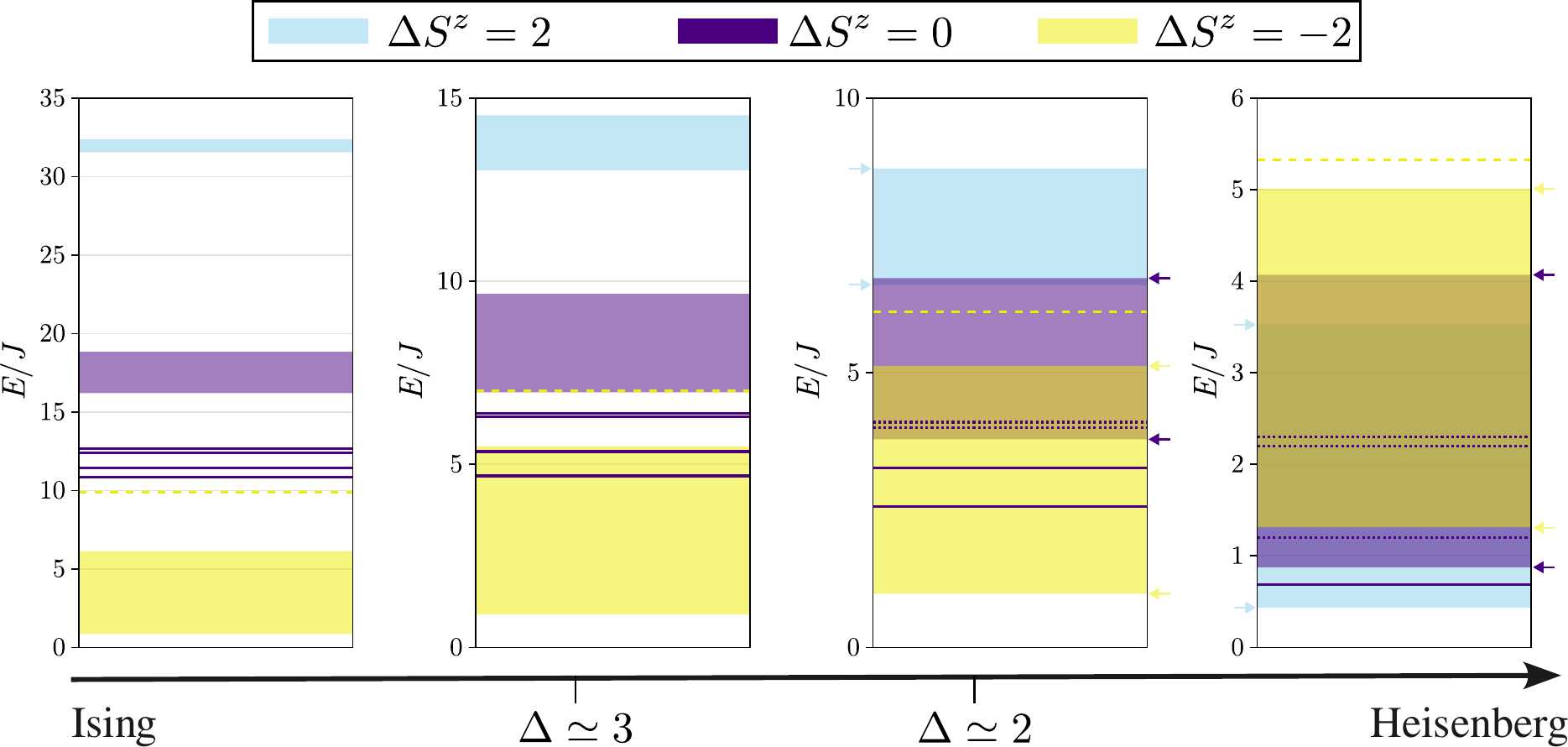}
    \caption{{\bf Schematic evolution of the two-magnon spectrum as a function of exchange anisotropy $\Delta$}. The color indicates the quantum number $\Delta S^z$ of the corresponding state. Shaded regions represent the two-magnon continua. For $\Delta S^z=0$, solid lines denote two-magnon bound states, while dotted lines indicate two-magnon resonances. Dashed lines represent quadrupolar bound states with $\Delta S^z=-2$. The arrows indicate the lower and upper bounds of the overlapping continua.}
    \label{fig:3}
\end{figure*}

As the system evolves from the Ising limit toward the isotropic point ($\Delta = 1$), two-magnon bound states become increasingly dispersive and eventually overlap with the continuum formed by pairs of single-magnon excitations. To capture this evolution, we adopt the approach of Ref.~\cite{XieT2023}, tracking the redistribution of spectral weight as a function of the exchange anisotropy $\Delta$. In contrast to previous studies, however, the THED method employed here offers a key advantage: it provides direct access to the many-body eigenstates within the truncated Hilbert space, enabling a more transparent identification of the underlying excitations.

Figure~\ref{fig:3} presents a schematic evolution of the two-magnon spectrum extracted from THED simulations as $\Delta$ is varied. Each two-magnon state is classified by its total spin change: $\Delta S^z = 2$ when both spins flip from ``down'' to ``up,'' $\Delta S^z = -2$ for two ``up'' to ``down'' flips, and $\Delta S^z = 0$ when one spin flips in each direction (color-coded in Fig.~\ref{fig:3}). 

Finite $xy$ interactions lift the degeneracy of the classical Ising energy $\delta E_c$ associated with the $\Delta S^z = 0$ bound state, splitting it into four distinct branches. The highest-energy branch corresponds to a state with lattice angular momentum $l = 0$ and energy $\delta E = \delta E_c + J$, while the lowest-energy branch has $l = \pi$ and $\delta E = \delta E_c - J$. The remaining two branches form quasi-doublets with angular momenta $l = \pm \pi/3$ and $l = \pm 2\pi/3$, located at $\delta E = \delta E_c + J/2$ and $\delta E = \delta E_c - J/2$, respectively. In the Ising limit, all four branches appear as sharp bound states, lying between the $\Delta S^z = -2$ and $\Delta S^z = 0$ continua.

As $\Delta$ decreases, the lower two branches are the first to enter the $\Delta S^z = -2$ continuum. However, their quantum numbers prohibit hybridization with the continuum, resulting in BICs. By contrast, the upper two branches merge into the $\Delta S^z = 0$ continuum at smaller $\Delta$, where hybridization is allowed, converting them into two-magnon resonances. At the Heisenberg point, the lower branches persist as BICs or as bound states below the continuum, while the upper branches remain as resonant features.

\begin{figure*}
    \centering
    \includegraphics[width=2\columnwidth]{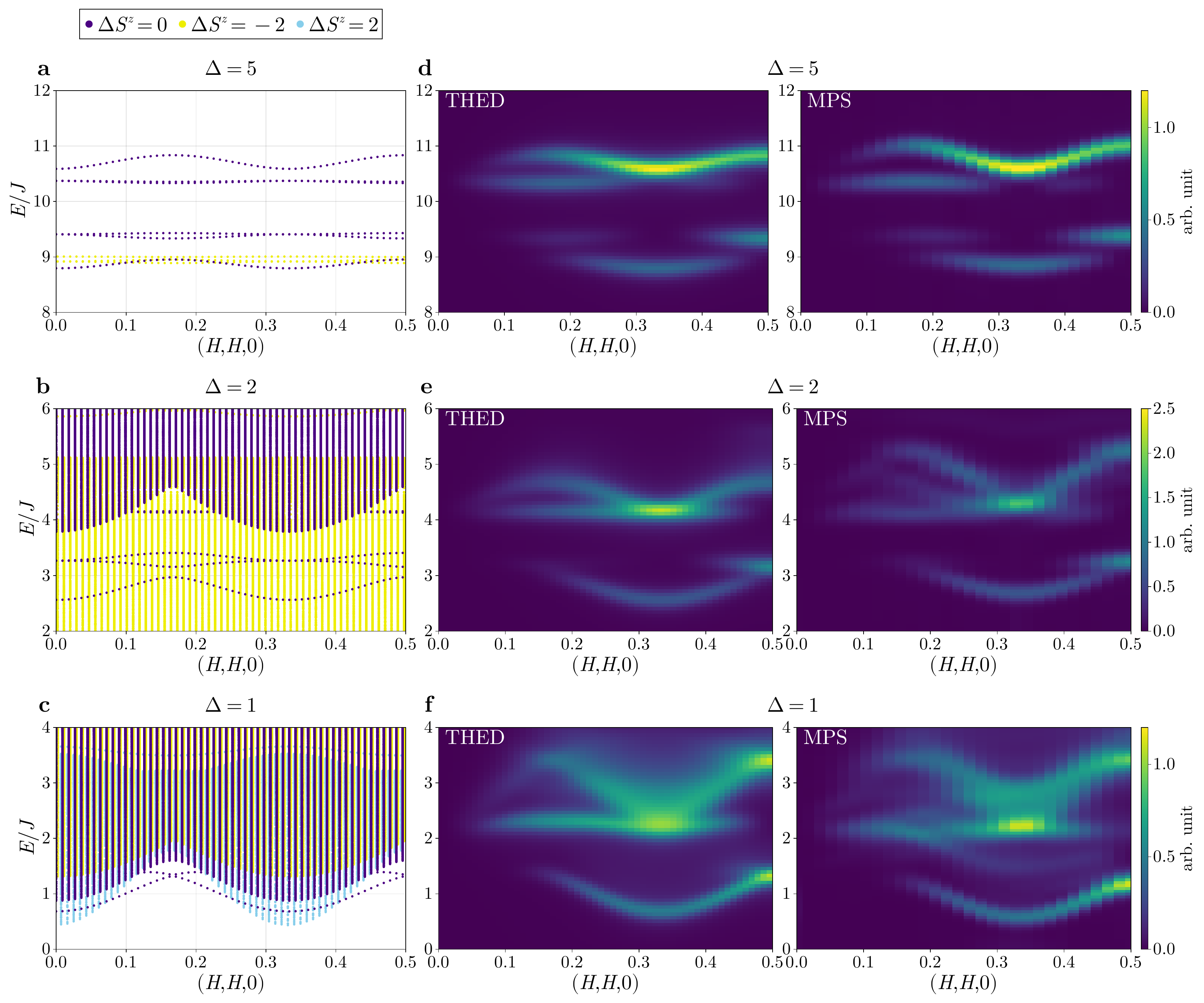}
    \caption{{\bf The two-magnon spectrum and the longitudinal dynamical spin structure factor $\mathcal{S}^{zz}(\bm{q},\omega)$ for different exchange anisotropies along the $(H,H,0)$ path.} {\bf a}--{\bf c} Two-magnon eigenvalues and quantum numbers $\Delta S^z$ obtained from the truncated Hilbert space exact diagonalization (THED) method for $\Delta = 5, 2$ and 1, respectively. {\bf d}--{\bf f} Comparison of $\mathcal{S}^{zz}(\bm{q},\omega)$ for the same values of $\Delta = 5, 2$, and 1, as computing using THED and the time evolution of the matrix product state (MPS).}
    \label{fig:4}
\end{figure*}

\begin{figure*}
    \centering
    \includegraphics[width=2\columnwidth]{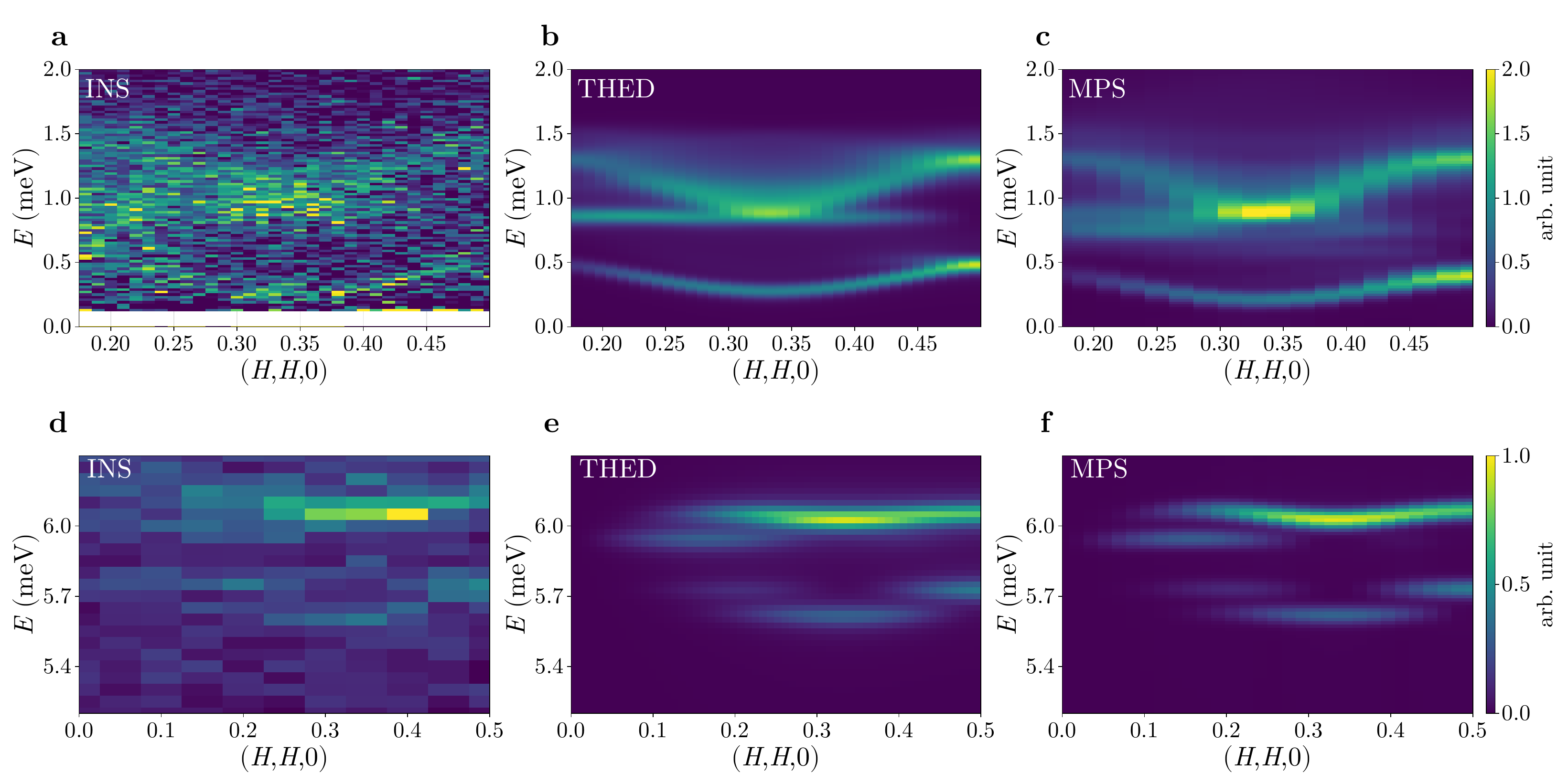}
    \caption{{\bf Additional experimental verifications}
    {\bf a, b, c} Comparison of the longitudinal dynamical spin structure factor $\mathcal{S}^{xx}(\bm{q},\omega)$ for $\mathrm{CsYbSe_2}$ at $B=4 \ T$ in the UUD phase obtained from INS, THED, and MPS.  {\bf d, e, f} Comparison of the INS intensities for $\mathrm{K_2Co(SeO_3)_2}$ with simulated results from THED and MPS.}
    \label{fig:5}
\end{figure*}

The schematic picture discussed above is based on identifying the lower and upper bounds of the continuum and isolated states from THED, neglecting their dispersion. The actual dispersions of the two-magnon eigenstates obtained from THED for $\Delta = 5$, $2$, and $1$ are shown in the first column of Fig.~\ref{fig:4} along the $[H,H,0]$ direction. In the two rightmost columns of Fig.~\ref{fig:4}, we present the calculated longitudinal dynamical spin structure factor (DSSF), defined as
\begin{equation}
    \mathcal{S}^{\alpha \alpha}(\bm{q},\omega) 
    = \frac{1}{2\pi N_s}\int_{-\infty}^{+\infty} dt 
    \sum_{j,l}^{N_s} e^{i\left(\omega t - \bm{q}\cdot(\bm{r}_j-\bm{r}_l) \right)}\langle S_{j}^{\alpha}(t) S_{l}^{\alpha}(0) \rangle,
\end{equation}
with $\alpha =z$ for the model Hamiltonian Eq.~\eqref{eq:model} with only nearest-neighbor exchange $J=J_1$ and $g\mu_B B/J_1 =1.905$, where $\bm{B}$ is applied along the $z$-direction and $N_s$ denotes the number of sites. These calculations are carried out using both THED and MPS methods. 
The longitudinal DSSF predicted here is experimentally accessible: it can be measured directly with polarized INS, or extracted from unpolarized INS provided the $g$-tensor is anisotropic, as illustrated by the examples in Fig.~\ref{fig:5}.

For $\Delta = 5$, we find  good agreement between the longitudinal dynamical structure factor $\mathcal{S}^{zz}(\bm{q},\omega)$ computed using THED and MPS. Both methods reveal four nearly flat and well-defined two-magnon branches along the $[H, H, 0]$ direction. A real-space analysis of the corresponding Fourier-transformed wavefunctions  obtained from THED and presented in the Supplementary Sec.~III, shows that given a spin flip on the ``up'' (``down'') sublattice, the second spin flip on the ``down'' (``up'') sublattice is strongly localized to the nearest-neighbor sites. This spatial confinement supports the identification of these modes as tightly bound two-magnon states.

For $\Delta = 2$, the lower two branches persist as BICs, in line with the discussion above, while the upper two evolve into resonances, appearing as broadened features in both THED and MPS spectra. Upon reaching the isotropic point $\Delta = 1$, parts of the lower branches detach from the continuum along the $[H, H, 0]$ direction, while others remain embedded. Nevertheless, due to their distinct quantum numbers, they retain their character as sharp BICs. The upper branches continue to manifest as resonances.

The real-space wavefunctions of these resonant states, also analyzed in Supplementary Sec.~III, exhibit extended spatial profiles: unlike bound states, whose wavefunction decays rapidly beyond a characteristic binding length, the resonance wavefunctions maintain significant amplitude at large separations, consistent with their delocalized nature.

Figure~\ref{fig:4} shows that the agreement between THED and MPS remains robust at the Heisenberg point. However, constant-$\bm{q}$ cuts, provided in Supplementary Figs.~S5, S6, and S7, highlight minor discrepancies between the two methods. First, as shown in Supplementary Figs.~S5 and S6, THED is not expected to capture the small renormalization of the single-magnon dispersion arising from hybridization with three-magnon states. This explains the slight mismatch in single-magnon peak positions at $\Delta = 1$, which could be resolved by extending the THED variational space to include three-magnon states. The smallness of this mismatch suggests that the maximum number of magnons provides a good control parameter for systematically expanding the low-energy DSSF.

For the two-magnon sector revealed by the longitudinal DSSF shown in Fig.~\ref{fig:4},  the MPS calculation reveals a weak intensity mode around $E/J = 1.5$ in the isotropic limit $\Delta = 1$, which does not appear in the THED results. This mode exhibits a parabolic dispersion, with its minimum at $H = 1/3$ coinciding with that of the two-magnon bound state near $E/J = 0.75$. This suggests that the additional mode seen in the MPS data may correspond either to a four-magnon bound state or to the onset of the two two-magnon bound-state continuum. Such an interpretation would also explain its absence in the present THED implementation, which includes only single and two-magnon states. However, as discussed in Supplementary Sec.~II, this mode could also be an artifact of the cylindrical geometries used in MPS.

Another discrepancy emerges at $\Delta = 2$ in the energy window $4.5 \lesssim E/J \lesssim 5.5$. The MPS results display a splitting between two modes for $H < 1/3$, which is absent in the THED calculation. 
Furthermore, while both approaches exhibit a single dominant high-intensity mode for $H > 1/3$, the mode identified by MPS appears at higher energy. The detailed analysis in Supplementary Sec.~II indicates that these discrepancies may once again stem from the proximity to the four-magnon continuum associated with the two-magnon bound state. At the same time, this analysis shows that finite-size effects due to the cylindrical geometry are pronounced in this region of the spectrum, suggesting that the MPS results may not fully capture the two-dimensional limit of the model.

The above discussion allows us to reinterpret the INS data for KYS. The structured continua observed in Fig.~\ref{fig:1}{\bf d} above the single-magnon branches originate from the spectral weight of two-magnon resonances, arising from the hybridization between two-magnon bound states and the two-magnon continuum. Due to their proximity to the single-magnon branches, the spectral weight of the two-magnon bound states overlap with that of the single-magnon modes, making them difficult to distinguish directly in INS (see Supplementary Fig.~S5 for detailed line cuts).

A key advantage of the Yb-based compound $\mathrm{CsYbSe_2}$ (CYS), described by $\mathcal{H}$ with $J_1 = 0.395~\text{meV}$ and $J_2/J_1 \simeq 0.03$, is the pronounced anisotropy of its $\bm{g}$-tensor ($g_{aa} = g_{bb} = 3.25$, $g_{cc} = 0.2$). This anisotropy enables the direct extraction of the longitudinal DSSF from INS intensities. In Fig.~\ref{fig:5}{\bf a}--{\bf c}, we compare the measured $\mathcal{S}^{xx}(\bm{q},\omega)$ with spins polarized along the $x$-axis at $B = 4~\mathrm{T}$ (deep inside the UUD phase) to theoretical results obtained from THED and MPS. The agreement corroborates our interpretation: sharp two-magnon bound states dominate the low-energy spectrum, while broader features at higher energies are consistent with two-magnon resonances.

Finally, we turn to the material $\mathrm{K_2Co(SeO_3)_2}$ (KCS) in its UUD phase, which realizes the model in Eq.~\eqref{eq:model} with nearest-neighbor interactions close to the Ising limit. The strong agreement between INS data and theoretical predictions from both THED and MPS simulations provides further validation of the two-magnon bound state picture. To reproduce the experimentally observed positions of the bound-state modes, we refine the exchange anisotropy from the previously reported value $\Delta = J_{zz}/J_{xy} = 14.28$~\cite{ZhuM2024a,ZhuM2024b} to $\Delta = 13.48$, while keeping $J_{xy} = 0.217\ \text{meV}$ fixed~\footnote{The original estimate of $\Delta$ was obtained in Ref.~\cite{ZhuM2024b} by fitting magnetization and specific heat using first-order perturbation theory in $J_{xy}$. The small discrepancy likely stems from higher-order corrections not accounted for in that analysis.}. This refinement underscores the utility of our approach in accurately extracting microscopic model parameters. Moreover, as shown in the Supplementary Fig.~S12, the absence of spectral weight associated with the quadrupolar bound state reinforces the conclusion that KCS is well described by an effective U(1)-symmetric Heisenberg model.

\section*{Discussion}

We have presented a comprehensive investigation of non-perturbative dynamics in the 1/3-plateau (UUD) phase of triangular lattice antiferromagnets. By combining two complementary numerical methods, truncated Hilbert space exact diagonalization and matrix product states, we demonstrate consistent and robust results, despite the distinct limitations of each approach: truncation errors in THED and finite-size effects in MPS. Benchmarking against inelastic neutron scattering experiments confirms the reliability of both methods for modeling quantum magnets in this regime.

The high expressiveness of MPS makes it a powerful tool for probing non-perturbative phenomena in two-dimensional ordered magnets, while the interpretability of THED allows us to disentangle the structure of anomalous continua and identify underlying bound-state features. 

Our results indicate that previously collected INS data on triangular lattice compounds in the UUD phase merit re-examination in light of this refined theoretical framework. Moreover, the role of higher-order $n_{\text{mag}}$-magnon states ($n_{\text{mag}} \geq 3$), omitted in the current THED implementation, underscores the need for methodological improvements, a feasible task given the polynomial scaling of the numerical complexity.  At the same time, efforts in advancing the MPS algorithms may enable simulations on wider cylinders, thereby reducing the finite-size effects introduced by the cylindrical geometry. Taken together, the combined use of THED and MPS provides a powerful and broadly applicable framework for studying non-perturbative excitations in ordered quantum magnets with gapped single-magnon spectra. Looking ahead, we anticipate that this dual approach will be especially valuable in contexts where conventional methods face limitations.

In this work, we focus on the canonical example of the triangular lattice in the UUD phase. More generally, the THED method is expected to be applicable to a broader class of gapped quantum magnets that exhibit semiclassical behavior. The two-magnon bound states and resonances discussed here can only be obtained through an infinite-order resummation in the $1/S$ expansion, where $S$ labels the irreducible representations (irreps) of SU(2). For systems with $S \geq 1$ or for coupled entangled units~\cite{Dahlbom24}, the classical limit of certain Hamiltonians, such as those with large single-ion anisotropy or weak interactions between entangled units, is more appropriately described by direct products of SU($N$) coherent states, where $N$ is the number of levels of each unit. 
In such cases, the $1/S$ expansion must be replaced by a $1/\lambda_1$ expansion, where $\lambda_1$ labels the completely symmetric irreps of SU($N$)~\cite{Muniz14,Hao2021,David2022,David2022b,David2023}. In terms for Feynman diagrams, the $1/\lambda_1$ expansion is simply the well-known loop expansion. In other words, the order of each diagram in powers of $1/\lambda_1$ is linear in  the number of loops.

The quadratic Hamiltonian obtained to order $\lambda_1$ corresponds to a generalized spin-wave theory that naturally incorporates different types of quasiparticles. For instance, these quasiparticles may correspond to triplon modes in weakly coupled dimer systems~\cite{Dahlbom24} or to dipolar and quadrupolar waves in $S=1$ systems~\cite{DoSH2021a,BaiX2021,LegrosA2021,BaiX2023}. Similar to the $1/S$ expansion, the formation of two-quasiparticle bound states requires an infinite resummation of ladder diagrams with an increasing number of loops. Both perturbative and non-perturbative effects have been demonstrated in effective $S=1$ ordered magnets~\cite{DoSH2021a,BaiX2021,LegrosA2021,BaiX2023}. In particular, higher-order bound states, such as four- and six-magnon states, have been shown to arise from non-perturbative effects in the excitation spectrum of FeI$_2$~\cite{LegrosA2021,BaiX2023}.  

To demonstrate the broader applicability of the THED approach, we have developed an open-source implementation~\cite{ZhangH2025} within the \texttt{Sunny.jl} spin-dynamics  framework~\cite{DahlbomD2025,DahlbomD2025a}. This framework provides a user-friendly interface for model definition and many-body calculations, following a workflow analogous to semiclassical linear spin-wave theory. Thanks to its flexibility and computational efficiency, THED offers a powerful tool for investigating gapped quantum magnets with semiclassical behavior in three dimensions, where MPS becomes impractical. Furthermore, as an efficient and accurate forward-scattering approach, THED can be seamlessly integrated into inverse scattering frameworks, particularly in light of recent advances in machine learning algorithms~\cite{samarakoon_2020}.

\bibliographystyle{apsrev4-2}
%

\section*{Acknowledgments}
We thank Shang-Shun Zhang, Kipton Barros, David Dahlbom, Martin Mourigal, Chaebin Kim, and Xiaojian Bai for useful discussions. H.Z. and C.D.B. thank Bruce Normand for his insightful comments and critical reading of the manuscript.
C.D.B. acknowledges support from the U.S. Department of Energy, Office of Science, Basic Energy Sciences, Materials Sciences and Engineering Division, under Award No. DE-SC-0018660. H.Z. acknowledges support from the Lincoln Chair of Excellence in Physics.
The work by A.S. is supported by the Quantum Science Center (QSC), a National Quantum Information Science Research Center of the U.S. Department of Energy (DOE).
Work at ETHZ was partially supported by a MINT grant of the Swiss National Science Foundation. Data on $\mathrm{K_2Co(SeO_3)_2}$ were collected using the AMATERAS spectrometer at
J-PARC in Experiment no. 2023B0161.
T. X. was supported by the National Key Research and Development Program of China (Grant No. 2024YFA1613100), the National Natural Science Foundation of China (Grant No. 12304187), the Guangzhou Basic and Applied Basic Research Funds (Grant No. 2024A04J4024), and the open research fund of Songshan Lake Materials Laboratory (Grant No. 2023SLABFN30).  The neutron scattering experiments of CsYbSe2 were performed on the time-of-flight (ToF) Cold Neutron Chopper Spectrometer (CNCS) at the Spallation Neutron Source at Oak Ridge National Laboratory (ORNL).

\section*{Author contributions}
C.D.B. and H.Z. conceived the project. H.Z. developed the numerical code for the THED calculations and analyzed the results together with C.D.B.. T.H. and A.M.L. carried out the MPS simulations. A.S. performed the INS experiments and analyzed the data for KYS. T.X. provided the INS data for CYS. M.Z. and A.Z. conducted the INS experiments and analyzed the data for KCS. N.M. and S.O.-K. assisted with neutron scattering experiments at J-PARC for KCS. H.Z., T.H., A.M.L., and C.D.B. wrote the manuscript with input from all authors.

\end{document}


\title{Supplementary Material For ``Nonperturbative Semiclassical Spin Dynamics for Ordered Quantum Magnets''}
\author{Hao~Zhang}
\affiliation{Department of Physics and Astronomy, The University of Tennessee,
Knoxville, Tennessee 37996, USA}
\author{Tianyue~Huang}
\affiliation{Institute of Physics, \'{E}cole Polytechnique F\'{e}d\'{e}rale de Lausanne (EPFL), CH-1015 Lausanne, Switzerland}
\author{Allen~O.~Scheie}
\affiliation{MPA-Q, Los Alamos National Laboratory, Los Alamos, New Mexico 87545, USA}
\author{Mengze~Zhu}
\affiliation{Laboratory for Solid State Physics, ETH Z\"{u}rich, 8093 Z\"{u}rich, Switzerland}
\author{Tao~Xie}
\affiliation{Center for Neutron Science and Technology, Guangdong Provincial Key Laboratory of Magnetoelectric Physics and Devices,
School of Physics, Sun Yat-sen University, Guangzhou, Guangdong 510275, China}
\author{N.~Murai}
\affiliation{J-PARC Center, Japan Atomic Energy Agency, Tokai, Ibaraki 319-1195, Japan}
\author{S.~Ohira-Kawamura}
\affiliation{J-PARC Center, Japan Atomic Energy Agency, Tokai, Ibaraki 319-1195, Japan}
\author{Andrey~Zheludev}
\affiliation{Laboratory for Solid State Physics, ETH Z\"{u}rich, 8093 Z\"{u}rich, Switzerland}
\author{Andreas~M.~L\"{a}uchli}
\affiliation{Laboratory for Theoretical and Computational Physics, Paul Scherrer Institut, CH-5232 Villigen-PSI, Switzerland}
\affiliation{Institute of Physics, \'{E}cole Polytechnique F\'{e}d\'{e}rale de Lausanne (EPFL), CH-1015 Lausanne, Switzerland}
\author{Cristian~D.~Batista}
\affiliation{Department of Physics and Astronomy, The University of Tennessee,
Knoxville, Tennessee 37996, USA}
\affiliation{Quantum Condensed Matter Division and Shull-Wollan Center, Oak Ridge
National Laboratory, Oak Ridge, Tennessee 37831, USA}
\date{\today}
\setcounter{figure}{0}
\global\long\def\thefigure{S\arabic{figure}}%
\setcounter{table}{0}
\global\long\def\thetable{S\arabic{table}}%
\setcounter{equation}{0}
\global\long\def\theequation{S\arabic{equation}}%

\maketitle

\section*{Methods}
\subsection{Truncated Hilbert Space Exact Diagonalization (THED)}
The truncated Hilbert space exact diagonalization (THED) method has previously been applied to the effective spin-one material $\mathrm{FeI_2}$, as described in Refs.~\cite{LegrosA2021,BaiX2023}. More recently, we have prototyped a general implementation~\cite{ZhangH2025} of this method within the framework of the open-source spin dynamics package \texttt{Sunny.jl}~\cite{DahlbomD2025,DahlbomD2025a}. Below, we briefly review the core aspects of the THED method and highlight specific details relevant to the calculations in the UUD phase.

We begin by noting that the UUD phase is the classical ground state of the Hamiltonian in Eq.~1 of the main text ($J_2=0$ and $\Delta_{ij}=1$) only for the fine-tuned external field value $B=3JS/(g\mu_B)$. Moreover, linear spin wave theory applied to this phase yields gapless single-magnon excitations at the ordering wave vector $\bm{q}=(1/3,1/3,0)$, despite the fact that the UUD order does not break any continuous symmetry. To resolve this inconsistency, we follow the approach introduced in Ref.~\cite{AliceaJ2009}, and diagonalize the renormalized quadratic Hamiltonian
\begin{equation}
    \tilde{\mathcal{H}}^{(2)}_{\text{UUD}} = \mathcal{H}^{(2)}_{\text{lswt}} + \mathcal{H}_{\text{nlswt}}^{(2)},
\end{equation}
where $\mathcal{H}^{(2)}_{\text{lswt}}$ is the linear spin wave theory Hamiltonian for the UUD phase, and $\mathcal{H}_{\text{nlswt}}^{(2)}$ captures the leading $1/S$ correction. The explicit forms of these terms can be found in Ref.~\cite{KamiyaY2018a}.  Importantly, the inclusion of $\mathcal{H}_{\text{nlswt}}^{(2)}$ opens a gap in the spin-wave spectrum, consistent with the fact that the UUD order does not break any continuous symmetry. The resulting quadratic Hamiltonian takes the form
\begin{equation}
    \tilde{\mathcal{H}}^{(2)}_{\text{UUD}} = \sum_{\bm{q}}\sum_{n=1}^3 
    E_{\bm{q},n}\left(\beta_{\bm{q},n}^{\dagger}\beta_{\bm{q},n} + \frac{1}{2} \right),
    \label{meq:uud_disp}
\end{equation}
where $\beta_{\bm{q},n}^{\dagger}$ creates a (gapped) single-magnon excitation at momentum $\bm{q}$ and band $n$ from the vacuum, denoted as $\lvert 0 \rangle$. 

To study interacting magnons and capture non-perturbative effects, we work within a truncated Hilbert space restricted to the 
$n_{\text{mag}}\leq2$ magnon subspace. This truncation is well justified due to the presence of a finite single-magnon gap. The basis of this subspace consists of single- and two-magnon states, denoted $\{ \lvert i \rangle, \lvert i \leq j \rangle \}$, where $|i\rangle = \beta_i^{\dagger} \lvert 0 \rangle$ and $\lvert i \leq j \rangle=\zeta_{ij}\beta_i^{\dagger} \beta_j^{\dagger}\lvert 0 \rangle$. Here $i$ labels the combined momentum $\bm{q}_i$ and band index $n_i$, while $\zeta_{ij}$ is the boson normalization factor defined as $\zeta_{i=j}=1/\sqrt{2!}$ and $\zeta_{i \neq j}=1$.

For the results presented this work, the THED calculations are performed on a finite triangular lattice comprising $37 \times 37$ magnetic unit cells (a total of 4107 spins). The full Hilbert space, of dimension $2^{4107}$, is truncated to a subspace containing at most two quasiparticles (magnons), reducing the dimensionality to $6165$  for each fixed center of mass momentum ($3$ single-magnon states plus $6162$ two-magnon states). Let $\mathcal{H}$ denote the original many-body spin Hamiltonian. The projected Hamiltonian in the truncated basis is then given by
\begin{equation}
    \mathcal{P}_{1,2} \mathcal{H} \mathcal{P}_{1,2} =
    \begin{pmatrix}
        \mathcal{H}_{11} & \mathcal{H}_{12} \\
        h.c. & \mathcal{H}_{22}
    \end{pmatrix},
    \label{meq:trun_ham}
\end{equation}
where the single-magnon block is diagonal:
\begin{equation}
    \mathcal{H}_{11}^{i,j}=\delta_{ij} E_{\bm{q}_i,n_i},
\end{equation}
and the two-magnon block is given by
\begin{equation}
    \mathcal{H}_{22}^{i \leq j, k \leq l}  = (\delta_{ik}\delta_{jl}+\delta_{il}\delta_{jk}) 
    (E_{\bm{q}_i,n_i}+E_{\bm{q}_j,n_j}) + \frac{1}{N_u} \tilde{U}_{n_i,n_j,n_k,n_l}^{(4)}(\bm{q}_i,\bm{q}_j,\bm{q}_k,\bm{q}_l)\zeta_{ij}\zeta_{kl}.
\end{equation}
The first term corresponds to the ``non-interacting'' two-magnon energy obtained in Eq.~\eqref{meq:uud_disp}, while the second encodes interactions via the four-magnon vertex. The full interaction Hamiltonian in the original magnon basis is
\begin{equation}
    \mathcal{H}_{2p}^{(4)}  =\frac{1}{2!2!N_{u}}\sum_{ijkl}\delta(\bm{q}_{i}+\bm{q}_{j}+\bm{q}_{k}+\bm{q}_{l}-\bm{G})
    \tilde{U}_{n_{i},n_{j},n_{k},n_{l}}^{(4)}(\bm{q}_{i},\bm{q}_{j},\bm{q}_{k},\bm{q}_{l})\beta_{\bar{\bm{q}}_{i},n_{i}}^{\dagger}\beta_{\bar{\bm{q}}_{j},n_{j}}^{\dagger}\beta_{\bm{q}_{k},n_{k}}\beta_{\bm{q}_{l},n_{l}},
    \label{meq:quartic_vtx}
\end{equation}
where $\bm{G}$ is a reciprocal lattice vector of the magnetic unit cell, ensuring momentum conservation within the first magnetic Brillouin zone. The four-magnon interaction vertex $\tilde{U}_{n_{i},n_{j},n_{k},n_{l}}^{(4)}(\bm{q}_{i},\bm{q}_{j},\bm{q}_{k},\bm{q}_{l})$ is expressed in terms of the generalized eigenvectors of the para-unitary (Bogoliubov) transformation that diagonalizes $\tilde{\mathcal{H}}^{(2)}_{\text{uud}}$.

In the UUD phase, the off-diagonal block vanishes, $\mathcal{H}^{12} = \mathcal{H}^{21} = 0$, as expected for generic collinear states in $S = 1/2$ magnets with U(1) symmetry~\cite{ChernyshevAL2009}. By contrast, for more general (non-collinear) magnetic orders, the off-diagonal terms associated with cubic interaction vertices remain finite and can induce decay and renormalization of the single-magnon modes~\cite{LegrosA2021,BaiX2023}.

To compare with matrix product state (MPS) and inelastic neutron scattering (INS) results, we compute the dynamical spin structure factor (DSSF) within the truncated Hilbert space restricted to $n_{\text{mag}} \leq 2$ magnons:
\begin{equation}
    \mathcal{S}^{\mu\mu}(\bm{q},\omega) = \sum_{\tilde{n}}
    \langle 0 \lvert S_{\bm{q}}^{\mu} \lvert \tilde{n} \rangle \langle \tilde{n} \lvert S_{-\bm{q}}^{\mu} \lvert 0\rangle \delta\left(\omega-(\mathcal{E}_{\bm{q},\tilde{n}}-\mathcal{E}_G)\right),
\label{eq:dssf_thed}
\end{equation}
where $\lvert \tilde{n} \rangle$ and $\mathcal{E}_{\tilde{n}}$ are the eigenstates and eigenvalues of the truncated Hamiltonian Eq.~\eqref{meq:trun_ham}, and $\mathcal{E}_G$ is the ground state energy. For the system size considered in this work (\textasciitilde 6000 basis states), all the eigenstates
$\lvert \tilde{n} \rangle$ can be obtained via full exact diagonalization. For larger systems, one may employ the continued fraction method introduced in Refs~\cite{GaglianoER1987,GaglianoER1988}, a Lanczos-based approach that circumvents the need for full diagonalization and reduce the numerical cost. In our case, we have verified that the continued fraction method yields results identical to exact diagonalization for a sufficiently large number of Lanczos iterations (typically > 1000).

\subsection{Matrix Product State Calculations}

\subsubsection{MPS spectral functions\label{seq:cal_S}}
\begin{figure}[t]
    \centering
    \includegraphics[width=0.75\linewidth]{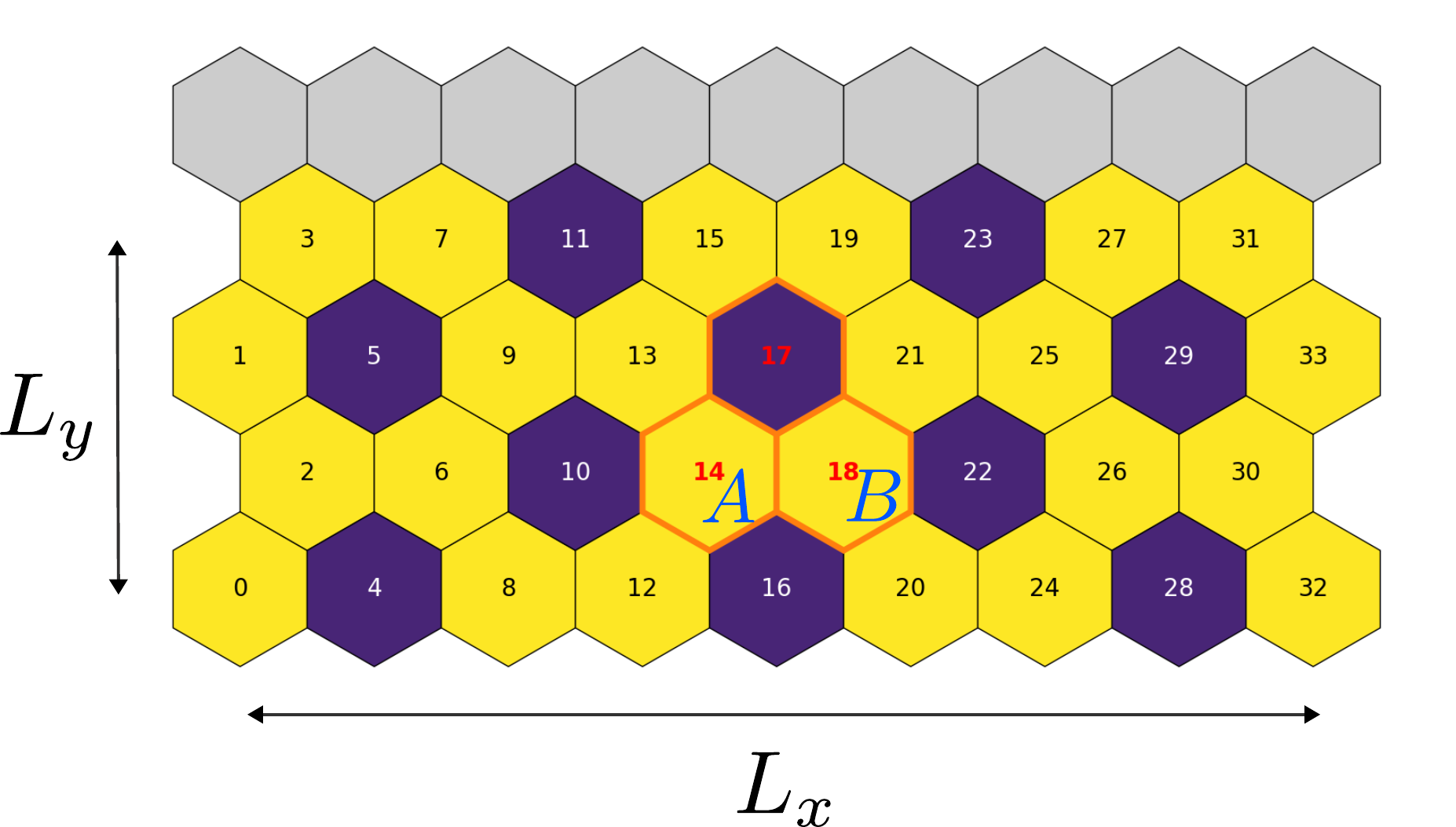}
    \caption{{\bf Illustration of the triangular lattice used in the matrix product state (MPS) calculations.} In this example, $L_x=9$ and $L_y=4$. Periodic boundary conditions are imposed along the $y$-direction - the grey row at the top represents the same row as at the bottom. The site numbers indicate their corresponding MPS indices, showing how the two-dimensional lattice is mapped onto a one-dimensional chain. The three sites marked out in the middle represent the central unit cell at which we apply the initial spin operator to start the time evolutions. We also fill all the sites with light and dark colors to represent the ``up'' and ``down'' spin states, respectively.}
    \label{fig:tri_lat}
\end{figure}
Matrix product states (MPS) provide an efficient variational ansatz for representing quantum many-body states in one-dimensional (1D) systems, where they are particularly effective for gapped ground states due to the area-law scaling of entanglement entropy \cite{schollwock_density-matrix_2011}. In quasi-two-dimensional (quasi-2D) geometries, such as cylinders of finite circumference, MPS can still yield accurate approximations provided that the entanglement growth along the mapped one-dimensional path remains moderate, typically requiring that the circumference not be too large.

In this work, we use MPS to compute the spectral function of a two-dimensional (2D) spin system. The 2D lattice is mapped onto a zig-zag path on a cylinder of circumference $L_y$ and length $L_x$, as illustrated in Fig.~\ref{fig:tri_lat}. For the data presented in the main text, we set $L_x = 30$ for the XXZ model in Fig.~4 and $L_x = 27$ for the three real compounds shown in Fig.~1 and Fig.~5, with $L_y = 6$ for all cases.  

The ground state is first obtained using the density-matrix renormalization group (DMRG). Time evolution is then simulated via the time-dependent variational principle (TDVP) algorithm, which projects the evolved states onto the MPS manifold~\cite{haegeman_time-dependent_2011, haegeman_unifying_2016, vanderstraeten_tangent-space_2019, paeckel_time-evolution_2019}. All calculations are carried out with the open-source Python library \texttt{TeNPy}~\cite{tenpy}. Exploiting the U(1) symmetry of the system, the time evolution is restricted to a fixed magnetization sector, which substantially reduces the computational cost.

In particular, we evaluate the connected part of the time-dependent correlation function
\begin{align}
    C^{\alpha\beta}(t, \bm{r}-\bm{r}_0) &= \bra{\Omega}S_{\bm{r}}^\alpha (t)S_{\bm{r}_0}^\beta(0)\ket{\Omega}_c \nonumber \\
    &= \bra{\Omega}S_{\bm{r}}^\alpha(t)S_{\bm{r}_0}^\beta(0)\ket{\Omega} - \bra{\Omega}S_{\bm{r}}^\alpha(t)\ket{\Omega}\bra{\Omega}S_{\bm{r}_0}^\beta(0)\ket{\Omega} \nonumber \\
    &= \bra{\Omega}S_{\bm{r}}^\alpha(t)S_{\bm{r}_0}^\beta(0)\ket{\Omega} - \bra{\Omega}S_{\bm{r}}^\alpha(0)\ket{\Omega}\bra{\Omega}S_{\bm{r}_0}^\beta(0)\ket{\Omega}
\end{align}
at each time step during the evolution. Here $\alpha, \beta \in \{+,-,z\}$, and $\ket{\Omega}$ denotes the ground state obtained from the DMRG calculation. Note that the $xx$ and $yy$ components of the correlation function can be obtained via $C^{xx}(t, \mathbf{r}-\mathbf{r}_0) = C^{yy}(t, \mathbf{r}-\mathbf{r}_0) = \frac{1}{4}[C^{+-}(t, \mathbf{r}-\mathbf{r}_0) + C^{-+}(t, \mathbf{r}-\mathbf{r}_0)]$. For all the results shown in the main text, we use $dt=0.05J_1^{-1} \sim 0.1J_1^{-1}$, and total time $T=30J_1^{-1} \sim 60J_1^{-1}$, where $J_1$ is the leading coupling strength.

Following the same rationale outlined in the supplementary material of Ref.~\cite{XieT2023}, we computed the desired correlation functions by time evolving three distinct sites at the central unit cell of the cylinder. Those sites correspond to the three sublattices of the symmetry broken UUD phase and are marked by red numbers in Fig.~\ref{fig:tri_lat}. This strategy significantly reduces the computational by allowing the use of a lower bond dimension (see Ref.~\cite{XieT2023} for details). 

Furthermore, we exploit the reflection symmetry of the UUD state to further reduce the number of independent simulations. Specifically, the correlation function originating from the two ``up'' sites--denoted as $A$ and $B$ (sites $14$ and $18$ in Fig.~\ref{fig:tri_lat}) are related by a reflection operation. By choosing a coordinate system in which sites $A$ and $B$ lie symmetrically along the $x$-axis with the origin at their midpoint, the following relation holds:
\begin{equation}
    C^{\alpha\beta}_A(t, x, y) = C^{\alpha\beta}_B(t, -x, y),
    \label{eq:c_reflection}
\end{equation}
where $C_A$ and $C_B$ are the correlation functions computed with the initial operator $S^{\beta}$ acting on site $A$ and site $B$, respectively. This symmetry reduces the number of required independent time evolutions from three to two, thereby lowering the computational cost of reconstructing the full spectral function. 

At the end of the calculation, we apply a Gaussian convolution function to the correlation function to suppress boundary effects from the finite cylinder size and artifacts from finite-time evolution:
\begin{equation}
    C^{\alpha\beta}(t, \bm{r}-\bm{r}_0) \longrightarrow \exp{-\frac{t^2}{2\sigma_t^2}}\exp{-\frac{\bm{r}^2}{2\sigma_r^2}}C^{\alpha\beta}(t, \bm{r}-\bm{r}_0)
\end{equation}
In our calculations, we use $\sigma_t = 14.4J_1^{-1}$ and $\sigma_r = 3.6a$ for most of the systems, where $a$ is the lattice constant. For the XXZ Hamiltonian with $\Delta=1$ used in Fig.~4 in the main text, an additional Gaussian time window with $\sigma_t \approx 7.07 J^{-1}$ applied.

After obtaining the time-dependent spin-spin correlation functions, we perform a Fourier transform to obtain the spectral function
\begin{equation}
    S^{\alpha\beta}_{\bm{r}_0}(\bm{q}, \omega) = \frac{1}{2\pi N_s}\int_{-T}^{T} \mathrm{d}t \sum_{\bm{r}} C^{\alpha\beta}(t, \bm{r}-\bm{r}_0) e^{i\left(\omega t - \bm{q}\cdot (\bm{r}-\bm{r}_0)\right)}.
\end{equation}
The correlation function at negative times, $C^{\alpha\beta}(-t, \bm{r})$, is obtained using time-reversal symmetry of the Hamiltonian, which ensures $C^{\alpha\beta}(-t, \bm{r}) = \overline{C^{\alpha\beta}(t, \bm{r})}$, where the overline indicates complex conjugation.

To obtain a spectral function that is directly comparable to experimental data, we compute an equal-weight average over the three initial sites within the central unit cell $u_c$:
\begin{equation}
    S^{\alpha\beta}_{\text{tot}}(\bm{q}, \omega) = \sum_{\bm{r}_0\in u_c} S^{\alpha\beta}_{\bm{r}_0}(\bm{q}, \omega).
    \label{eq:MPS_corr}
\end{equation}
where $u_c$ stands for the central unit cell. Furthermore, using the reflection symmetry in Eq.~\ref{eq:c_reflection}, one can show that
\begin{equation}
     S^{\alpha\beta}_A(q_x, q_y, \omega) = S^{\alpha\beta}_B(-q_x, q_y, \omega),
\end{equation}
which allows us to calculate $S^{\alpha\beta}_{\text{tot}}(\bm{q}, \omega)$ from the two independent time evolutions we have simulated.

\begin{figure}[h]
    \centering
    \includegraphics[width=0.85\linewidth]{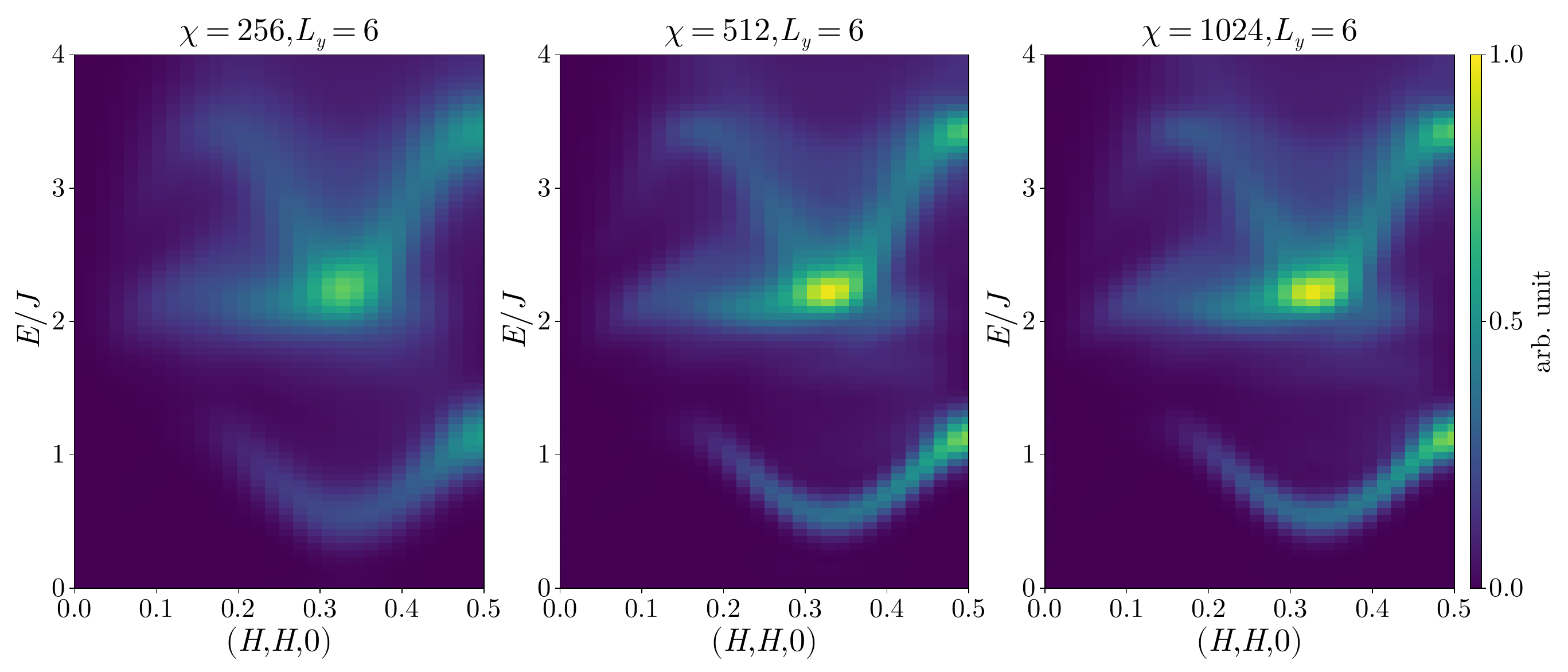}
    \caption{Comparison of spectral functions computed with bond dimensions $\chi=256$, $512$ and $1024$ for the Heisenberg model $\Delta=1$ on triangular lattice for $L_y=6$.}
    \label{fig:bond_dim}
\end{figure}

Finally, we verify the convergence of our results with respect to the bond dimension. As shown in Fig.~\ref{fig:bond_dim}, increasing the bond dimension from $\chi=512$ to $\chi=1024$ for systems with $L_y=6$ results in only minor changes, indicating that $\chi=512$ is sufficient for reliable results in the cases studied.

\begin{figure}[h]
    \centering
    \includegraphics[width=0.85\linewidth]{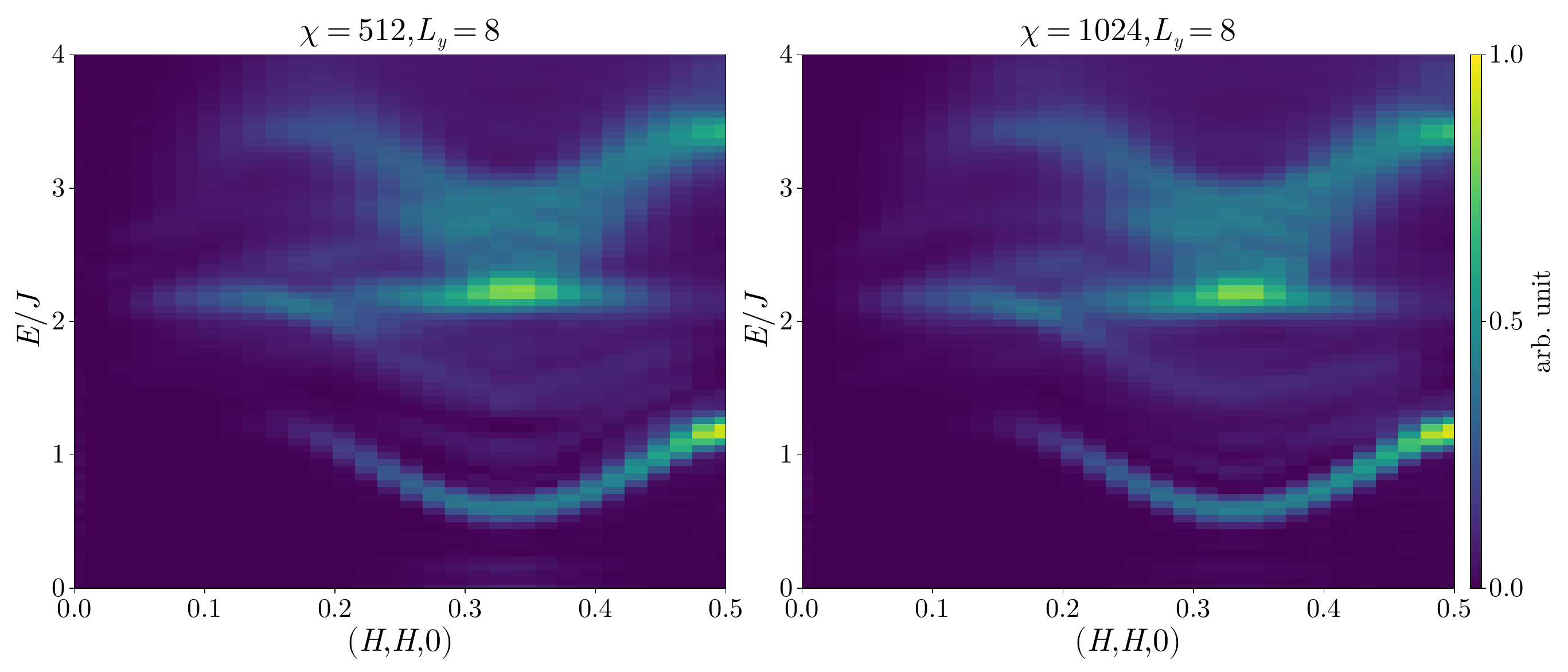}
    \caption{Comparison of spectral functions computed with bond dimensions $\chi=512$ and $1024$ for the triangular Heisenberg model ($\Delta=1$) with $L_y=8$.}
    \label{fig:bond_dim_8}
\end{figure}

Therefore, all simulations presented in Figs.~1 and 5 of the main text were performed with a bond dimension of $\chi=512$, except for $\kcs$, where a smaller value of $\chi=64$ was sufficient. This reduction is justified by the system's proximity to the Ising limit, which leads to lower entanglement entropy and thus reduced computational cost. For the XXZ model results shown in Fig.~4 of the main text, we employed a cylinder of size $L_x=27, L_y=8$ with bond dimension $\chi=1024$ in order to obtain a clearer understanding of the discrepancies between the two numerical methods, as discussed below. The corresponding convergence with respect to $\chi=1024$ for $L_y=8$ is shown in Fig.~\ref{fig:bond_dim_8}.

\subsubsection{Post processing: removing the unphysical recurrence of Bragg peaks}
\begin{figure}
    \centering
    \includegraphics[width=1.0\linewidth]{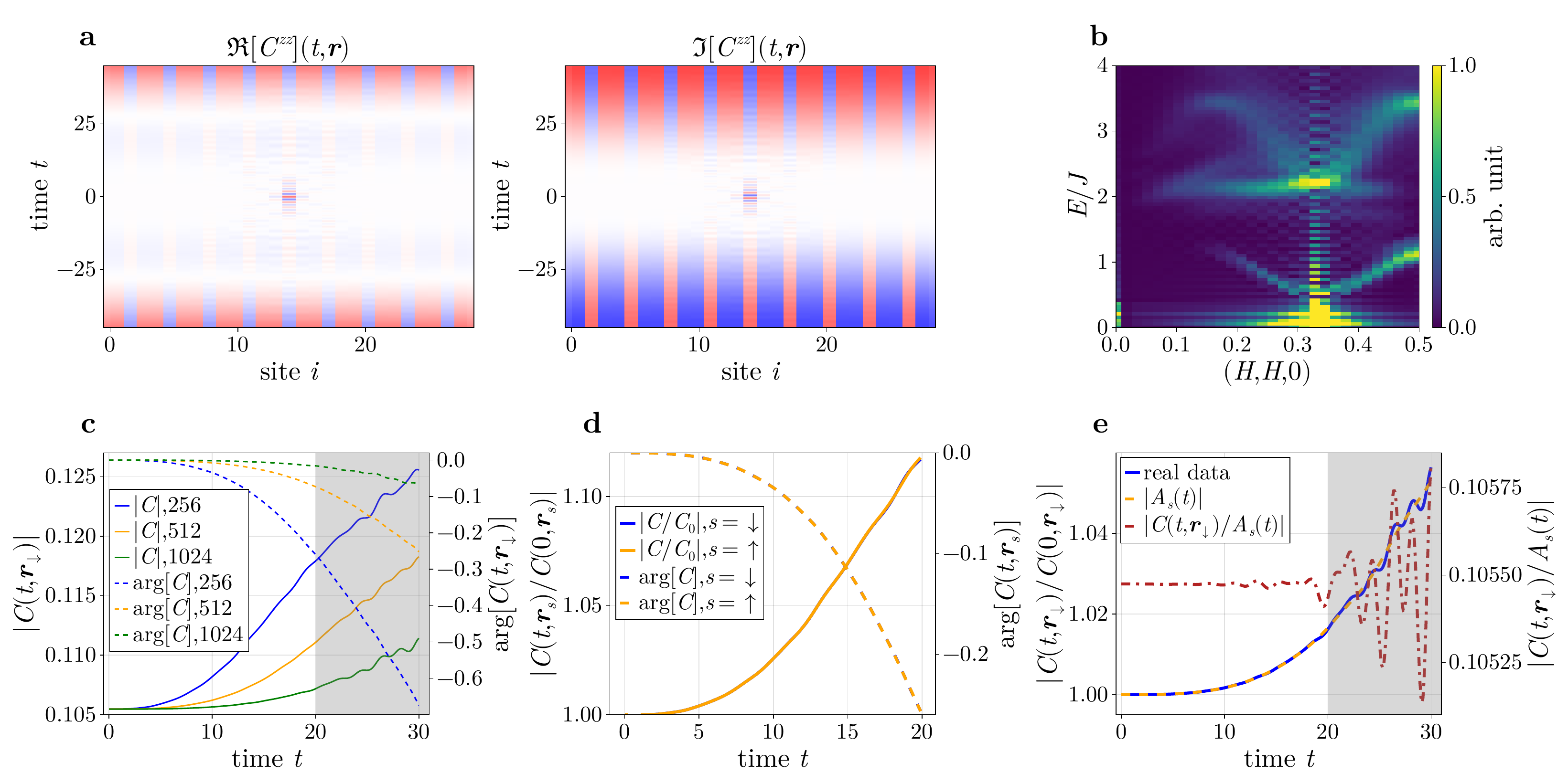}
    \caption{(a) The original connected time-dependent spin spin correlation function for Heisenberg model $\Delta=1$ with $\chi=256$. It shows $C(t, \bm{r})$ of sites in the same row as the initial excited site. (b) The corresponding spectral function with the same Gaussian convolution function applied as before. Artificial intensity around $\Gamma$ and $K$ points hindered us in reading off the information of the spectra. (c) The dependence of correlation functions on the bond dimensions for the same Heisenberg model. Solid lines stand for the norm of the correlations and dashed lines show the phase shifts. (d) The values of $C(t,\bm{r})/C(0,\bm{r})$ for a Up-site and a Down-site near the boundary. They overlap quite perfectly. (e) The extracted amplification curve from real simulation data, and the post-processed correlation function is shown in red.}
    \label{fig:MPS_corr}
\end{figure}

During our calculations, we found that in the UUD phase, the connected time-dependent correlation function $C^{zz}(t, \mathbf{r})$ exhibits an unphysical recurrence of magnetization order, which becomes increasingly amplified over time--even though the disconnected part, $\langle S^z_\mathbf{r} \rangle \langle S^z_{\mathbf{r}_0} \rangle$, has been subtracted. This artifact results in an artificially enhanced intensity near the Bragg peaks in the corresponding spectral function (Fig.~\ref{fig:MPS_corr}{\bf a} and Fig.~\ref{fig:MPS_corr}{\bf b}), potentially obscuring the finer spectral features we aim to resolve.

Evidence for the artificial nature of this recurrence is twofold.  
First, its spatial profile violates the light-cone structure imposed by the Lieb-Robinson bound: the apparent order emerges simultaneously and uniformly across the entire system, rather than propagating ``causally'' from the excitation center. Moreover, the recurrence is independent of the position at which the initial operator $S^z$ is applied.  
Second, the effect shows a strong dependence on the bond dimension, as illustrated in Fig.~\ref{fig:MPS_corr}{\bf c}. To quantify this, we examine $|C(t, \bm{r}=\bm{r}_{\text{down}})|$, where $\bm{r}_{\text{down}}$ denotes a spin down-site located far from the excitation center. For all bond dimensions considered, $|C(t, \bm{r}=\bm{r}_{\text{down}})|$ exhibits an artificial growth over time, accompanied by a phase shift away from $0$. However, both the amplitude enhancement and phase distortion diminish systematically as the bond dimension $\chi$ increases, and we expect these artifacts to vanish in the $\chi \to \infty$ limit. This strongly suggests that the recurrence is a numerical artifact rather than a genuine physical effect.  
Finally, we note that this issue is not specific to the TDVP algorithm used for time evolution: a similar artifact has been reported for time-evolving block decimation (TEBD), as shown in Fig.~22 of Ref.~\cite{paeckel_time-evolution_2019}. This points to the MPS structure itself, rather than any particular time-evolution scheme, as the likely source of the artifact.

To systematically characterize this recurrence, we define an \emph{amplification curve},  
\[
A_s(t) = \frac{C(t, \bm{r} = \bm{r}_s)}{C(0, \bm{r} = \bm{r}_s)},
\]  
where $\bm{r}_s$ denotes the position of a site located far from the initially excited center (chosen near the middle of the system), such that within the relevant time window it remains outside the light cone imposed by the Lieb-Robinson bound. The normalization by $C(0, \bm{r} = \bm{r}_s)$ removes the trivial difference between spin-up and spin-down sites. Consequently, the profile of $A_s(t)$ directly reflects how the correlation function deviates from its initial value. The solid line in Fig.~\ref{fig:MPS_corr}{\bf d} shows an example of $A_s(t)$ obtained from our simulations, clearly indicating that $A_s(t)$ is independent of the initial spin state at the chosen site $\bm{r}_s$.

Based on our observations, we assume that all sites in the system are affected by the same amplification factor $A_s(t)$. This allows us to perform a post-processing correction by undoing the artificial amplification,  
\[
C^{zz}(t, \mathbf{r}) \;\longrightarrow\; \frac{C^{zz}(t, \mathbf{r})}{A_s(t)},
\]  
directly on the data obtained from time-evolution algorithms such as TDVP.
 

Nevertheless, a practical subtlety arises: because we aim to evolve the system for as long as possible, and due to the exponential correlation tails outside the light cone, we cannot directly obtain the ideal amplification curve over the entire time window. As shown in Fig.~\ref{fig:MPS_corr}{\bf e}, for $t \gtrsim 20$ $A_s(t)$ begins to fluctuate and increase non-monotonically, which we attribute to the approaching light cone. Consequently, we approximate the true amplification curve $A_s(t)$ by fitting a smoothing spline to the measured value. The orange dashed line in Fig.~\ref{fig:MPS_corr}{\bf e} illustrates an example of this procedure.

In conclusion, we removed the spurious recurrence of magnetization order observed during time evolution by dividing the obtained correlation functions by the extracted amplification curve $A_s(t)$. This procedure, described in Sec.~\ref{seq:cal_S}, yields a significantly cleaner spectral function. All $S^{zz}(\bm{q}, \omega)$ data presented in the main text and in Sec.~\ref{seq:cal_S} of this supplementary material were obtained after applying this post-processing correction.

\section{Constant-\texorpdfstring{$\bm{q}$}{q} Cuts of Intensity Data}
In this section, we present constant-$\bm{q}$ cuts of the intensity data obtained from inelastic neutron scattering (INS) experiments, as well as from truncated Hilbert space exact diagonalization (THED) and matrix product state (MPS) calculations, as shown in Figs.~1, 4, and 5 of the main text. These detailed comparisons allow for a clearer visualization of the agreement among the data and facilitate a more precise analysis of their discrepancies as discussed in Sec.~\ref{sec:discre}.

\subsection{Materials: \texorpdfstring{$\kys$}{KYbSe2}, \texorpdfstring{$\cys$}{CsYbSe2}, and \texorpdfstring{$\kcs$}{K2Co(SeO3)2}}
To enable a quantitative comparison with the INS data of $\kys$, we apply both the neutron polarization factor and the $\bm{g}$-tensor specific to $\kys$ to the dynamical spin structure factor (DSSF) computed using THED [Eq.~\eqref{eq:dssf_thed}] and MPS [Eq.~\eqref{eq:MPS_corr}]. The resulting INS intensity is given by
\begin{equation}
    I(\bm{q},\omega) = \sum_{\mu=x,y,z} \left(1- \frac{q^{\mu}q^{\mu}}{\lvert \bm{q} \rvert^2} \right)(g^{\mu \mu})^2\mathcal{S}^{\mu\mu}(\bm{q},\omega) ,
\end{equation}
where $g^{xx} = g^{yy} = 3.02$ and $g^{zz} = 1.86$ are the components of the $\bm{g}$-tensor reported in Ref.~\cite{ScheieAO2024}, and $\mathcal{S}^{\mu\mu}(\bm{q},\omega)$ denotes the diagonal components of the DSSF.
Due to experimental challenges, the determination of the $g$-factors carries significant uncertainties~\cite{ScheieAO2024}. In our THED and MPS calculations, we fix the Zeeman energy as $g^{xx} \mu_B B = 1.73J_1$ ($J_1=0.456$ meV and $\alpha$ is the field direction)  to simulate the INS data at $B = 4 \ \mathrm{T}$, and subsequently apply the $g$-tensor corrections to the calculated DSSF. We emphasize that only these diagonal terms contribute to the INS intensity for the present system.

The resulting comparison among INS, THED, and MPS data is presented as heatmaps in Fig.~1{\bf d} of the main text. Here we focus on three representative $\bm{q}$ points along the $[q_1, q_1, 3]$ direction, specifically at $q_1 = 0.15$, $1/3$, and $0.5$. To highlight the role of magnon-magnon interactions, we also include results from the non-interacting two-magnon continuum, computed using Eq.~\eqref{meq:uud_disp} with the interaction vertex $\tilde{U}$ in Eq.~\eqref{meq:quartic_vtx} set to zero.

\begin{figure}
    \centering
    \includegraphics[width=.85\linewidth]{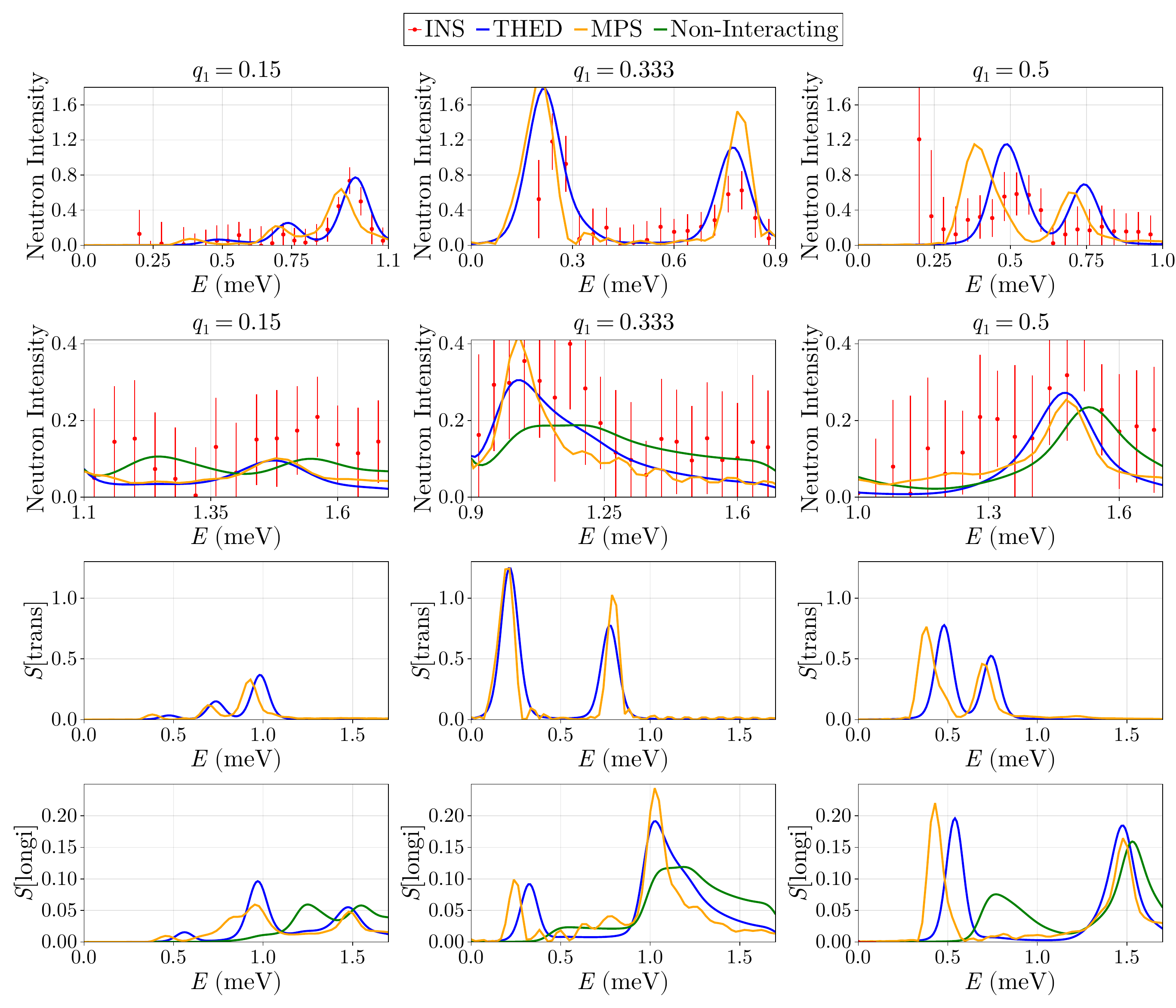}
    \caption{Comparison of constant-$\bm{q}$ cuts for $\kys$ at three representative momenta along the $[q_1,q_1,3]$ direction. 
    {\bf Rows 1--2:} Inelastic neutron scattering (INS) intensity versus theoretical results. Row 1 highlights the low-energy sector dominated by single-magnon contributions, compared with truncated Hilbert space exact diagonalization (THED) and matrix product state (MPS) calculations. 
    Row 2 focuses on the high-energy sector dominated by two-magnon contributions, compared with THED, MPS, and the non-interacting two-magnon continuum. 
   {\bf Row 3:} Transverse dynamical spin structure factors from THED and MPS. 
   {\bf Row 4:} Longitudinal dynamical spin structure factors from THED, MPS, and the non-interacting two-magnon continuum.}
    \label{fig:kys_cuts}
\end{figure}

As shown in the first two rows of Fig.~\ref{fig:kys_cuts}, both THED and MPS closely reproduce the INS data for $\kys$, capturing the peak positions and widths with high fidelity. To disentangle the different contributions to the DSSF, we show the transverse channel, which probes single-magnon excitations, in the third row, and the longitudinal channel, which captures two-magnon contributions in the fourth row. For consistent quantitative comparison, the non-interacting, THED, and MPS data in Fig.~\ref{fig:kys_cuts} and Fig.~1\textbf{d} of the main text have been convolved with a Gaussian resolution function of width $\sigma = 0.1J_1^{\text{kys}}$.

Minor discrepancies between THED and MPS arise in the single-magnon peaks at $q_1 = 0.15$ and $0.5$, likely due to three-magnon contributions not included in the THED framework. Notably, at $q_1 = 1/3$, where the single-magnon mode is well separated from the three-magnon continuum (see Sec.~\ref{sec:discre} for further discussion), the agreement between the two methods is excellent.

The longitudinal spectra in the fourth row of Fig.~\ref{fig:kys_cuts} exhibit similar consistency between THED and MPS, with only minor deviations arising from the same truncation limitations in THED. Importantly, both methods reveal marked deviations from the non-interacting two-magnon continuum, highlighting the substantial role of magnon-magnon interactions in reshaping the spectral lineshape. This provides clear evidence of pronounced non-perturbative effects in $\kys$.

We note that the two-magnon bound state appears as a weak, low-lying feature--visible as the first peak in all three panels of the bottom row--located in close proximity to the single-magnon peak. Its low intensity and overlap with nearby excitations render it challenging to resolve unambiguously in the INS data.

\begin{figure}
    \centering
    \includegraphics[width=0.85\linewidth]{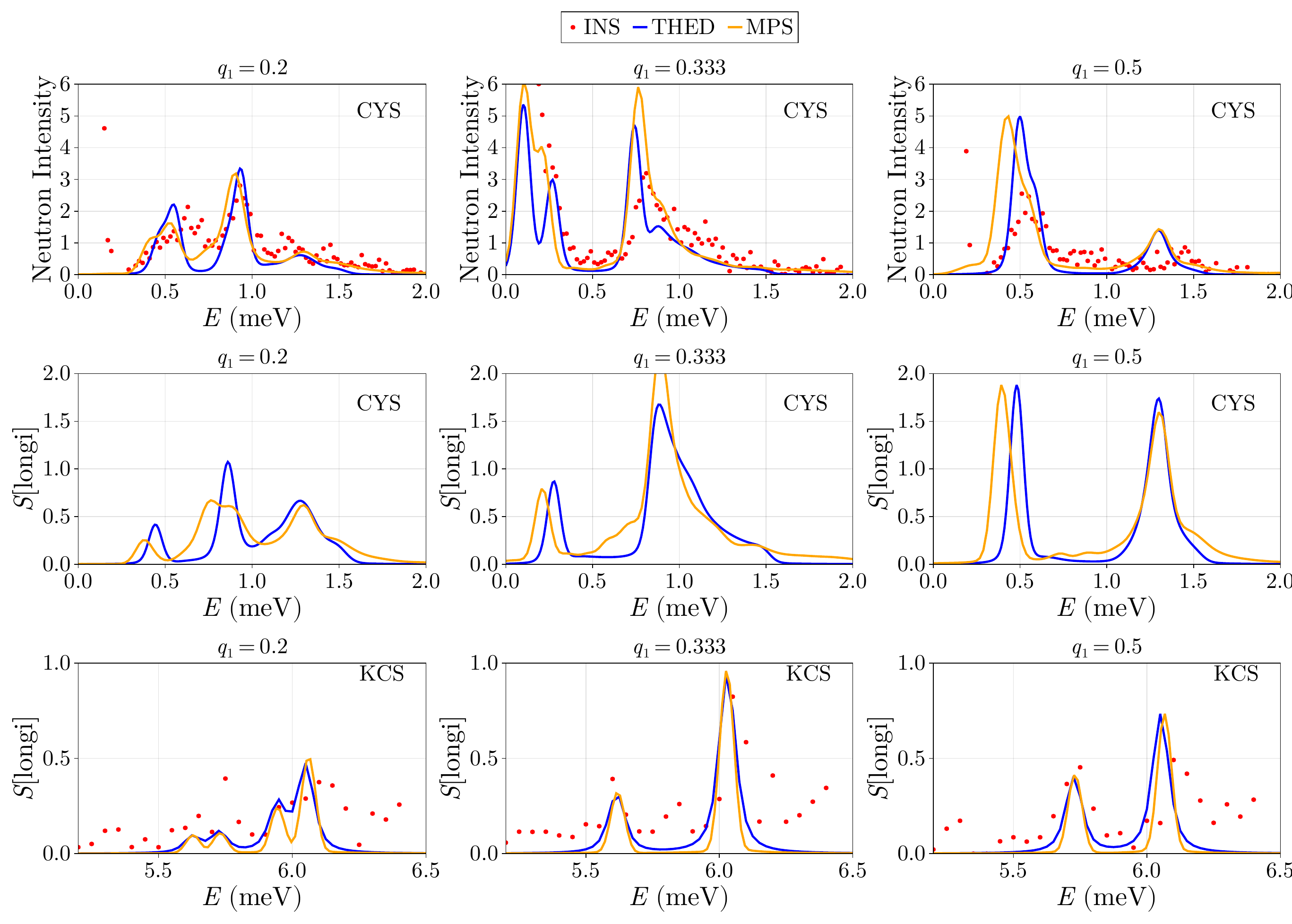}
    \caption{Comparison of constant-$\bm{q}$ cuts for $\cys$ and $\kcs$. Top row: INS data compared with simulated results from THED and MPS along the momentum path $[q_1, q_1, 2.5]$. Middle row: Longitudinal component of the dynamical spin structure factor for $\cys$. Bottom row: Longitudinal component of the dynamical spin structure factor for $\kcs$.}
    \label{fig:add_cuts}
\end{figure}

Figure~\ref{fig:add_cuts} presents constant-$\bm{q}$ cuts of the intensity data for both $\cys$ and $\kcs$. To enable quantitative comparison, the simulated data shown here and in Fig.~5 of the main text have been convoluted with a Gaussian wave packet of width $\sigma = 0.1J_1$, where $J_1$ denotes the nearest-neighbor interaction for the corresponding compound. Due to the strong anisotropy in the $g$-tensor of these materials--$g_{ab} = 3.25$ and $g_{cc} = 0.2$ for $\cys$ (with in-plane UUD order), and $g_{ab} = 2$, $g_{cc} = 7.9$ for $\kcs$ (with UUD order along the $c$-axis)--the longitudinal component of the DSSF can be effectively extracted from the INS data. As discussed in the main text, for materials with a weak easy-plane exchange anisotropy, we neglect this small contribution and model the system using a pure Heisenberg interaction. This approximation preserves the $\mathrm{U}(1)$ symmetry of $\mathcal{H}$ for arbitrary field orientations.

In the top row of Fig.~\ref{fig:add_cuts}, both THED and MPS show good quantitative agreement with the INS spectra for $\cys$. In the middle row, we display only the simulated longitudinal DSSF, as the corresponding component extracted from the INS data is too noisy to allow a meaningful comparison. A clearer illustration of the agreement between simulation and experiment can be found in the heatmap shown in Fig.~5 of the main text.

Finally, in the bottom row of Fig.~\ref{fig:add_cuts}, THED and MPS results exhibit excellent agreement for $\kcs$, which lies deep in the Ising regime with $\Delta = J_{zz}/J_{xy} = 14.28$, further validating the accuracy and consistency of both theoretical approaches in this limit.

\subsection{XXZ Hamiltonian}
\label{sec:xxz}
\begin{figure}
    \centering
    \includegraphics[width=0.85\linewidth]{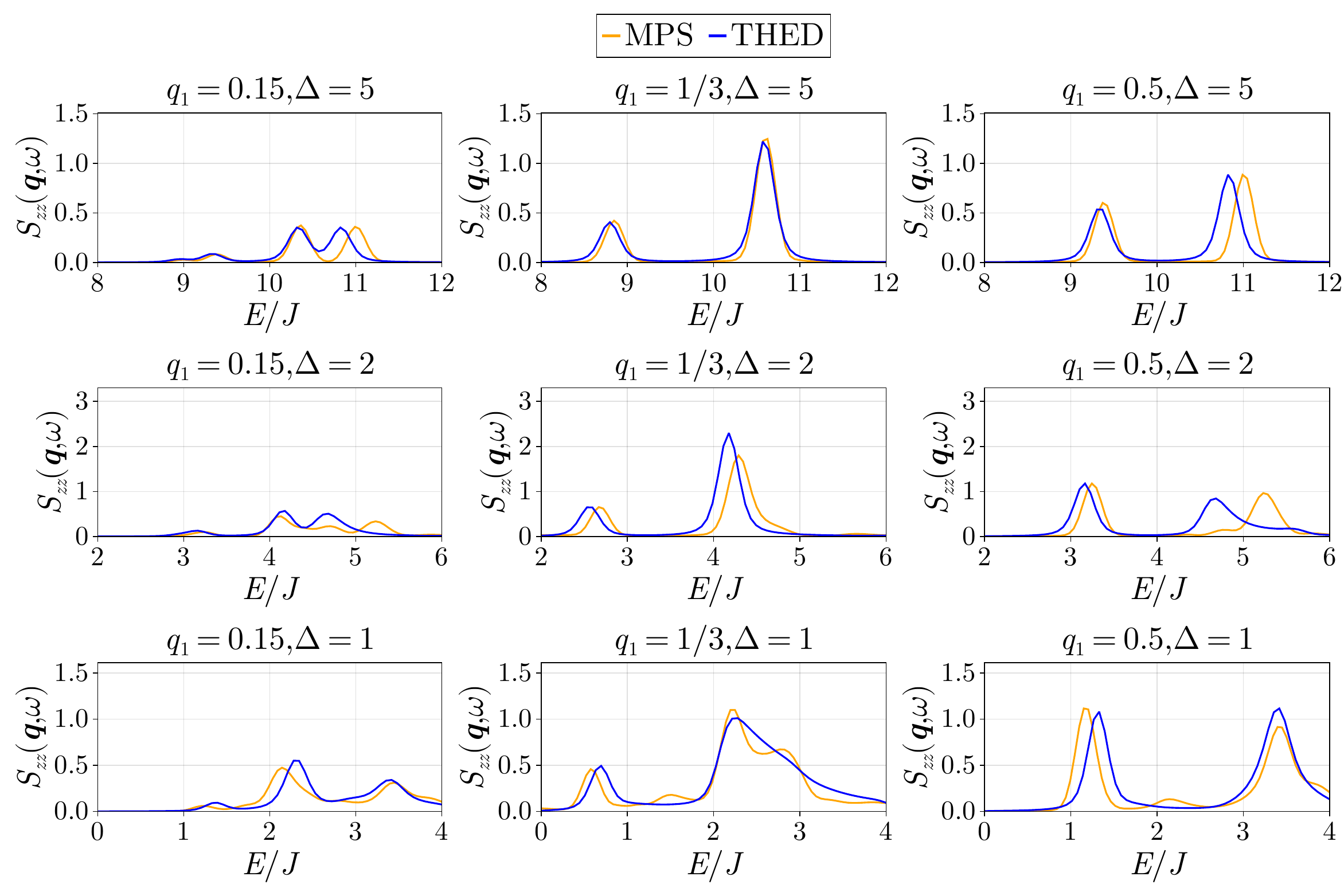}
    \caption{Comparison of constant-$\bm{q}$ cuts of the longitudinal dynamical spin structure factor for the XXZ model at different anisotropy parameters $\Delta$. THED results are shown alongside MPS data.}
    \label{fig:xxz_cuts}
\end{figure}

Figure~\ref{fig:xxz_cuts} presents a comparison of constant-$\bm{q}$ cuts of the longitudinal DSSF for the XXZ model at three representative anisotropy values: $\Delta = 5$, $2$, and $1$. As with the other data sets, both the THED and MPS results are convolved with a Gaussian resolution function of width $\sigma = 0.1J$. These results correspond to those shown in Fig.~4 of the main text. The MPS data were obtained by time-evolving a cylindrical system of size $L_x=27, L_y=8$ with bond dimension $\chi=1024$. The rationale for using a wider cylinder for the XXZ Hamiltonian is provided in Sec.~\ref{sec:discre}. Across the entire range, from the Ising limit to the Heisenberg point, we find good agreement between the two methods, with only minor discrepancies, which are discussed in detail below.

\section{Analysis of discrepancies}
\label{sec:discre}
In this section, we analyze the potential sources of discrepancies between the THED and MPS results, which may arise from limitations inherent to the current implementations of both methods. In particular, the truncation of $n_{\text{mag}} \geq 3$ magnon states in THED and the use of cylindrical geometries in MPS calculations represent the primary sources of deviation.

\subsection{Truncation of multi-magnon states}
\begin{figure}
    \centering
    \includegraphics[width=0.85\linewidth]{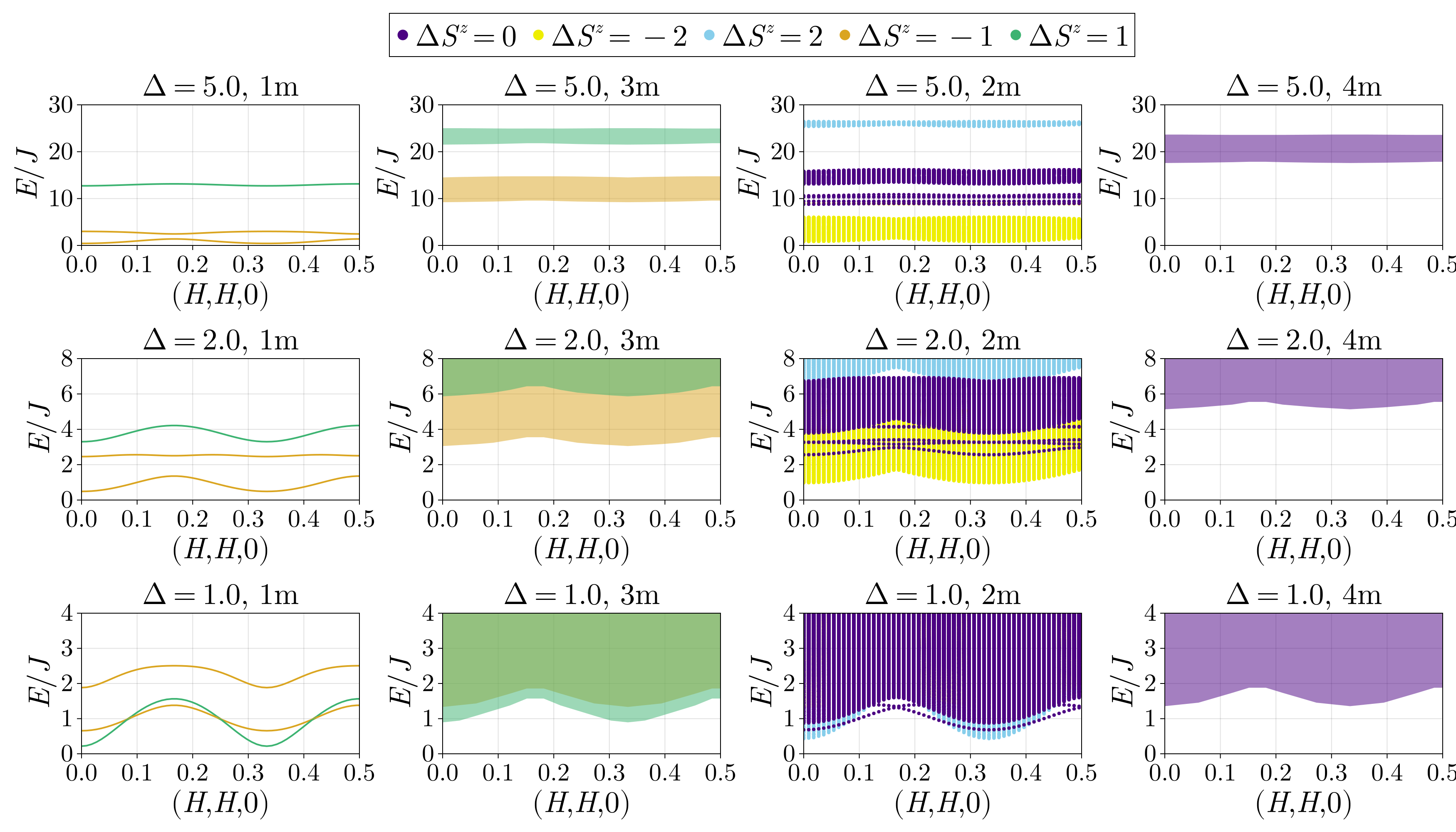}
    \caption{ One- to four-magnon states estimated from THED. The first two columns show the one- and three-magnon states, while the last two columns show the two- and four-magnon states. Different rows correspond to exchange anisotropies $\Delta = 5, 2, 1$. The one- and two-magnon states are obtained directly from THED up to two-magnon states, whereas the three- and four-magnon continua are constructed from the one-magnon states and the lowest two-magnon bound state band, and from the two two-magnon bound states from the lowest band, respectively. The color indicates the quantum number $\Delta S^z$ of the corresponding state.} 
    \label{fig:1234magnon}
\end{figure}

To estimate the contributions of $n_{\text{mag}} \geq 3$ states, Fig.~\ref{fig:1234magnon} displays the one-, three-, two-, and four-magnon states in four columns, with different rows corresponding to exchange anisotropies $\Delta = 5, 2, 1$. The one- and two-magnon states are obtained directly from THED up to two-magnon states, whereas the three- and four-magnon continua are constructed from the one-magnon states and the lowest two-magnon bound state band, and from the two two-magnon bound states from the lowest band, respectively.

If the Hilbert space is extended to include contributions up to four magnons, then, in addition to the particle-number conserving term $\beta^{\dagger}\beta^{\dagger}\beta\beta$ already incorporated in the current THED implementation, one must also account for particle-number non-conserving quartic interaction vertices of the form $U^{(4)}_{\text{p-non}} \sim \beta^{\dagger}\beta\beta\beta$. These vertices connect one- to three-magnon states and two- to four-magnon states, provided the sectors share the same quantum number $\Delta S^z$ (color-coded in Fig.~\ref{fig:1234magnon}). Since the model under consideration lacks a cubic vertex, no matrix elements exist between sectors with different magnon-number parity (e.g. between one- and two-magnon states or between one- and four-magnon states).

The small but systematic downshift of the single-magnon peak positions at the Heisenberg point $\Delta = 1$ can be attributed to $U^{(4)}_{\text{p-non}}$ through level repulsion. As shown in the third row of Fig.~\ref{fig:kys_cuts} and the first row of Fig.~\ref{fig:add_cuts}, the downshift is smaller for $q_1 = 1/3$ compared to $q_1 = 0.15$ and $q_1 = 0.5$, consistent with the fact that $q_1=1/3$ lies further from the three-magnon continuum (see Fig.~\ref{fig:1234magnon}).

Moreover,  as shown in Fig.~\ref{fig:xxz_cuts}, interactions between two- and four-magnon states mediated by $U^{(4)}_{\text{p-non}}$ may also contribute to the discrepancies between the longitudinal DSSF obtained from THED and MPS. Specifically, the MPS calculation reveals a weak intensity mode in the energy window $ 1.5\lesssim E/J \lesssim 2$ at $\Delta = 1$ along the $[q_1,q_1,0]$ direction for $q_1\simeq 0.2-0.5$, which is absent in the THED results. As indicated in Fig.~\ref{fig:1234magnon}, this mode coincides with the lower edge of the estimated four-magnon continuum formed by two two-magnon bound states. This suggests that the additional feature in MPS may correspond either to a four-magnon bound state or to the onset of the two two-magnon bound state continuum. Such an interpretation would also explain its absence in the current THED framework, which includes only single- and two-magnon states. However, as shown in Fig.~\ref{fig:MPS_finite_size}, this mode is not observed in MPS calculations for $L_y = 6$, while a similar feature does emerge in THED calculations for cylindrical geometries with small aspect ratios (see Fig.~\ref{fig:thed_finite_size}). Consequently, the discrepancy could also be attributed to finite size effects.

At $\Delta=2$, discrepancies in the longitudinal DSSF arise in the energy range $ 4.5  \lesssim E/J \lesssim 5.5 $. In this window, the MPS results exhibit a splitting between two modes for $H<1/3$, whereas THED shows only a single branch. For $H>1/3$, both approaches yield a dominant high-intensity mode, but the MPS mode lies at systematically higher energy. Again, as shown in Fig.~\ref{fig:1234magnon}, this interval coincides with the overlap of two-magnon states and the four-magnon continuum, highlighting the role of higher-magnon sectors in accounting for these differences.

The overall good agreement between THED and MPS indicates that the contribution of $U^{(4)}_{\text{p-non}}$ is minor for this problem. As discussed below, the discrepancies in the longitudinal DSSF shown in Fig.~\ref{fig:xxz_cuts} may also arise from the cylindrical geometry employed in MPS. Extending the current THED implementation to include the four-magnon sector increases the computational cost only from linear to cubic in system size, making the incorporation of three- and four-magnon states a promising and tractable avenue for future work to further reduce truncation errors and clarify the origin of these discrepancies.

\subsection{The impact of cylindrical geometry}
\begin{figure}
    \centering
    \includegraphics[width=0.85\linewidth]{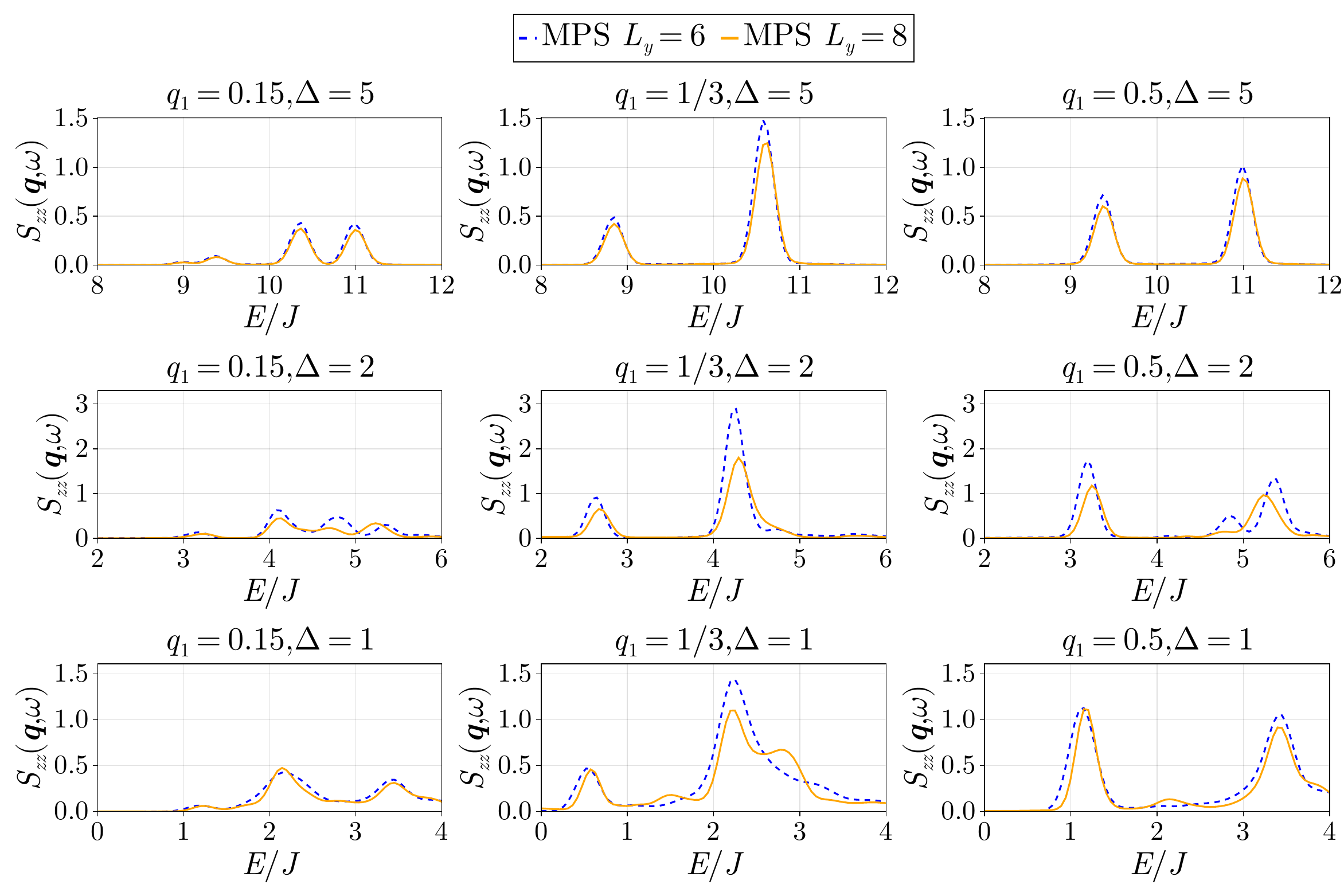}
    \caption{Longitudinal dynamical spin structure factor at constant-$\bm{q}$ from MPS simulations with $L_y=6$ and $L_y=8$ for three exchange anisotropies.}
    \label{fig:MPS_finite_size}
\end{figure}

Our MPS calculations are carried out on cylindrical geometries, with open boundaries along the $x$ direction of length $L_x$ and periodic boundaries along the $y$ circumference of length $L_y$. A key assumption is that the entanglement growth along the effective one-dimensional mapping remains manageable, which typically requires a moderate circumference. To examine the impact of geometry, we computed the longitudinal DSSF of the XXZ model for cylinders with $L_y=6$ and $L_y=8$. The convergence with respect to bond dimension is shown in Fig.~\ref{fig:bond_dim} for $L_y=6$ ($\chi=512$) and Fig.~\ref{fig:bond_dim_8} for $L_y=8$ ($\chi=1024$). A direct comparison of the longitudinal DSSF for the two cylinder widths is given in Fig.~\ref{fig:MPS_finite_size}, which reveals that even this modest increase in circumference leads to noticeable shifts in peak positions and splittings at certain wave vectors.

\begin{figure}
    \centering
    \includegraphics[width=0.85\linewidth]{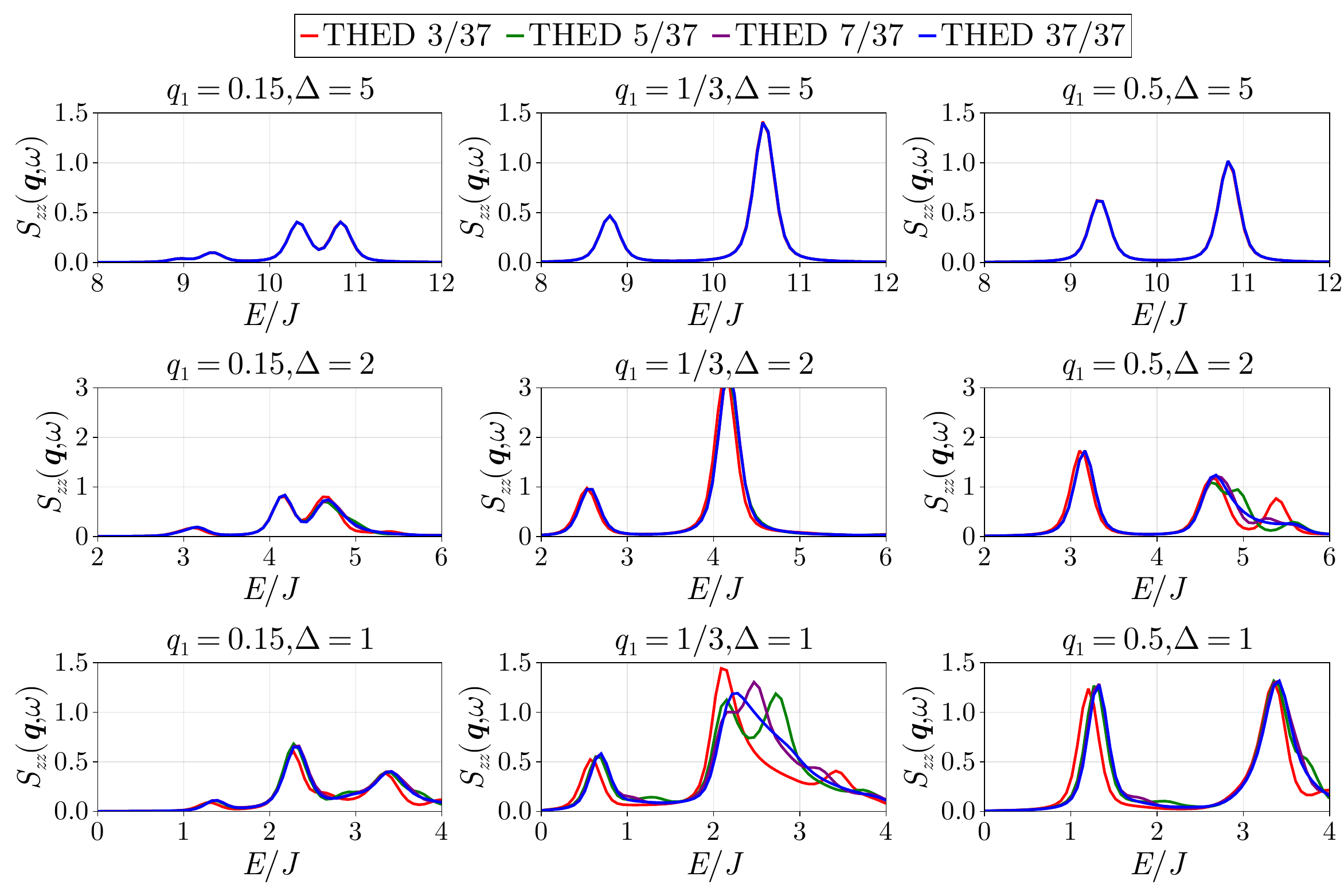}
    \caption{THED calculations in a cylinder geometry with different aspect ratios of magnetic unit cells along two directions with periodic boundary conditions.}
    \label{fig:thed_finite_size}
\end{figure}

To further assess finite-size effects, we performed additional THED calculations on clusters with varying aspect ratios, all with periodic boundary conditions in both directions. We find that the longitudinal DSSF $\mathcal{S}^{zz}$ is highly sensitive to the aspect ratio, particularly at $\Delta = 2$ and $\Delta = 1$. As shown in Fig.~\ref{fig:thed_finite_size}, the cylindrical geometry can induce spectral-weight splittings in these cases, coinciding with the largest discrepancies between the two methods--for example, at $\Delta = 2$, $q_1 = 0.5$, and $\Delta = 1$, $q_1 = 1/3$. Calculations with smaller aspect ratios exhibit spurious splittings compared to results closer to the thermodynamic limit.

In summary, while some of the discrepancies between THED and MPS can be attributed to the truncation of $n_{\text{mag}} \geq 3$ states in THED, they may also originate from the cylindrical geometry used in MPS. These findings highlight the need for methodological improvements in both approaches to resolve such ambiguities. Overall, the detailed comparison presented here reinforces that the combined use of THED and MPS provides a powerful framework for understanding non-perturbative spin dynamics in ordered quantum magnets.

\section{Interpretability and real space wave function}
\begin{figure}
    \centering
    \includegraphics[width=0.9\linewidth]{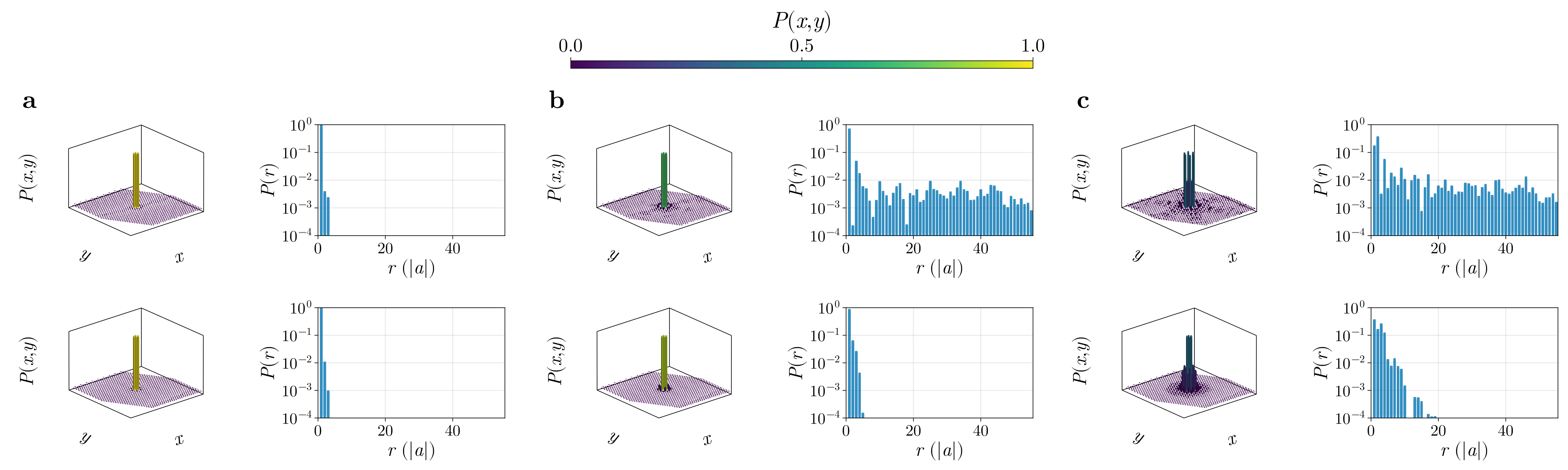}
    \caption{Squared real-space wave functions for the two states with the highest spectral intensity at center-of-mass momentum $\bm{q} = [1/3, 1/3, 0]$ shown in Fig.~4 of the main text. The top row corresponds to the state with the highest intensity on the upper branch, while the bottom row shows results for the state on the lower branch. {\bf a} ($\Delta=5)$, {\bf b} ($\Delta=2$), {\bf c} ($\Delta=1$): The first column displays the spatial distributions of the squared wave functions in linear coordinates. The second column shows angle-integrated results on a semi-logarithmic scale to highlight the asymptotic behavior.}
    \label{fig:real_wf}
\end{figure}
We now demonstrate that the THED method offers valuable interpretability through direct access to the eigenstates of the truncated Hamiltonian. Let us begin with the unnormalized two-spin-flip wave function
\begin{equation}
    \psi(\bm{0},\bm{r}; \alpha, \beta) = b_{\bm{0}\alpha}^{\dagger} b_{\bm{r}\beta}^{\dagger} \lvert 0 \rangle,
\end{equation}
which represents the probability amplitude for creating one spin flip at the origin (magnetic unit cell $\bm{0}$) on sublattice $\alpha$ and another at site $\bm{r}$ on sublattice $\beta$. Here, $b^{\dagger}_{\alpha}$ creates the Holstein-Primakoff boson in the local frame of the $\alpha$th sublattice, and $\lvert 0 \rangle$ is the vacuum of the Bogoliubov quasiparticles $\beta^{\dagger}$ (see Eq.~\eqref{meq:uud_disp}).

The eigenstates from exact diagonalization of the truncated Hamiltonian are labeled by center-of-mass momentum $\bm{q}$ and band index $\tilde{n}$, i.e., $\lvert \bm{q}, \tilde{n} \rangle$. We compute the squared overlap between this eigenstate and the real-space wave function:
\begin{equation}
    \lvert \langle \bm{q}, \tilde{n} \lvert \psi(\bm{0},\bm{r}; \alpha, \beta)\rangle \rvert^2
    = \bigg\vert\frac{1}{N_{u}}\sum_{\bm{k}\prime}\frac{U_{\tilde{n}}^{*}(\bm{q},\bm{k})}{\zeta(\bm{q}-\bm{k},n_{1};\bm{k},n_{2})}\bigg[e^{-i(\bm{q}-\bm{k})\cdot\bm{r}}[V_{\bm{k}}]_{\alpha,n_{2}}^{*}[V_{\bm{q}-\bm{k}}]_{\beta,n_{1}}^{*}+e^{-i\bm{k}\cdot\bm{r}}[V_{\bm{q}-\bm{k}}]_{\alpha,n_{1}}^{*}[V_{\bm{k}}]_{\beta,n_{2}}^{*}\bigg]\bigg\vert^{2},
    \label{meq:wavefunction}
\end{equation}
where $U_{\tilde{n}}(\bm{q},\bm{k})$ is the eigenvector of the truncated Hamiltonian corresponding to band $\tilde{n}$, and $[V_{\bm{k}}]_{\alpha,n_1}$ denotes the Bogoliubov transformation matrix element that para-diagonalizes Eq.~\eqref{meq:uud_disp}, with $\alpha,\beta$ labeling the sublattice and $n_1,n_2$ the single-magnon band index. The prime on the momentum summation indicates that it is restricted to two-particle states with fixed center-of-mass momentum $\bm{q}$.

In Fig.~\ref{fig:real_wf}, we plot the (normalized) squared wave functions obtained from Eq.~\eqref{meq:wavefunction} for the two states with the highest spectral weight at $\bm{q} = (1/3, 1/3, 0)$, as shown in Fig.~4 of the main text. For bound states, the probability of finding the second spin flip decays exponentially beyond a characteristic binding length. In contrast, for resonant states, this probability remains finite even at large separations, giving rise to long tails in the distribution.

This distinction is made clearer in the angle-integrated plots shown on a semi-logarithmic scale. The bound states exhibit clear exponential decay within the finite cluster, while the resonance states display extended profiles, indicating delocalization. These observations reinforce the physical picture that THED captures both bound and resonance states, offering a powerful framework for interpreting features in the dynamical spin structure factor.

Direct access to the real-space wavefunction allows us to explore an instructive scenario: we begin by creating a spin flip at the origin, $\bm{r}_{10} = \bm{0}$, on sublattice $\alpha_0$, followed by a second spin flip at a neighboring site, $\bm{r}_{20} = \bm{r}_{10} + \bm{r}_0$ (with $\bm{r}_0 = \bm{a}$), on sublattice $\beta_0$. This initial configuration is not an eigenstate of the truncated Hamiltonian and thus evolves under its dynamics.

To probe this evolution, we evaluate the probability of finding the second spin flip at a relative displacement $\bm{r} = \bm{r}_2 - \bm{r}_1$ with respect to the first, at a given time $t$. This is computed through the transition amplitude at fixed center-of-mass momentum $\bm{q} = (1/3, 1/3, 0)$:

\begin{align}
    C(\bm{r}, t; \bm{r}_{0}, \alpha, \beta, \alpha_{0}, \beta_{0})
    &= \langle \psi(\bm{r}; \alpha, \beta) | e^{-i \tilde{H}_{\bm{q}} t} | \psi(\bm{r}_{0}; \alpha_{0}, \beta_{0}) \rangle
    \nonumber \\
    & = \sum_{\tilde{n}}e^{-i\mathcal{E}_{\bm{q},\tilde{n}}t}\langle\psi(\bm{r};\alpha,\beta)\lvert\bm{q},\tilde{n}\rangle\langle\bm{q},\tilde{n}\lvert\psi(\bm{r}_{0};\alpha_{0},\beta_{0})\rangle
    \label{meq:transition}
\end{align}
where $\tilde{H}_{\bm{q}}$ denotes the truncated Hamiltonian introduced in Eq.~\eqref{meq:trun_ham}, evaluated at $\bm{q} = (1/3, 1/3, 0)$, and $\mathcal{E}_{\bm{q},\tilde{n}}$ labels the eigenvalues of $\tilde{H}_{\bm{q}}$.

In the videos submitted with this article ``\textit{dynamical\_wave\_function\_$\Delta$\_(5.0,1.0).mp4},'' we visualize the time evolution of the quantum state $\psi(\bm{r}_0; \alpha_0, \beta_0)$ by plotting the square of the absolute value of the transition amplitude defined in Eq.~\eqref{meq:transition} (after normalization). In the UUD phase, the Ising limit ($\Delta = 5$) clearly exhibits two spin flips that remain tightly bound over time, with occasional quantum tunneling of the second spin flip to a different nearest-neighbor site.

In contrast, at the Heisenberg point ($\Delta = 1$), the two spin flips form a transient bound pair that persists over a finite lifetime $\tau \sim \hbar / |\Gamma|$, where $\Gamma$ is the average width (over momentum) of the resonance in $\mathcal{S}^{\mu\mu}(\bm{q},\omega)$ . Beyond this time scale, the pair decays into two independent magnons.

We anticipate that in situ correlation measurements, such as those proposed in Ref.~\cite{FukuharaT2013}, could directly detect this real-space, time-resolved evolution of the quantum states, as illustrated in the simulations and accompanying video files.

\section{U(1) symmetry breaking term}
\begin{figure}
    \centering
    \includegraphics[width=0.85\linewidth]{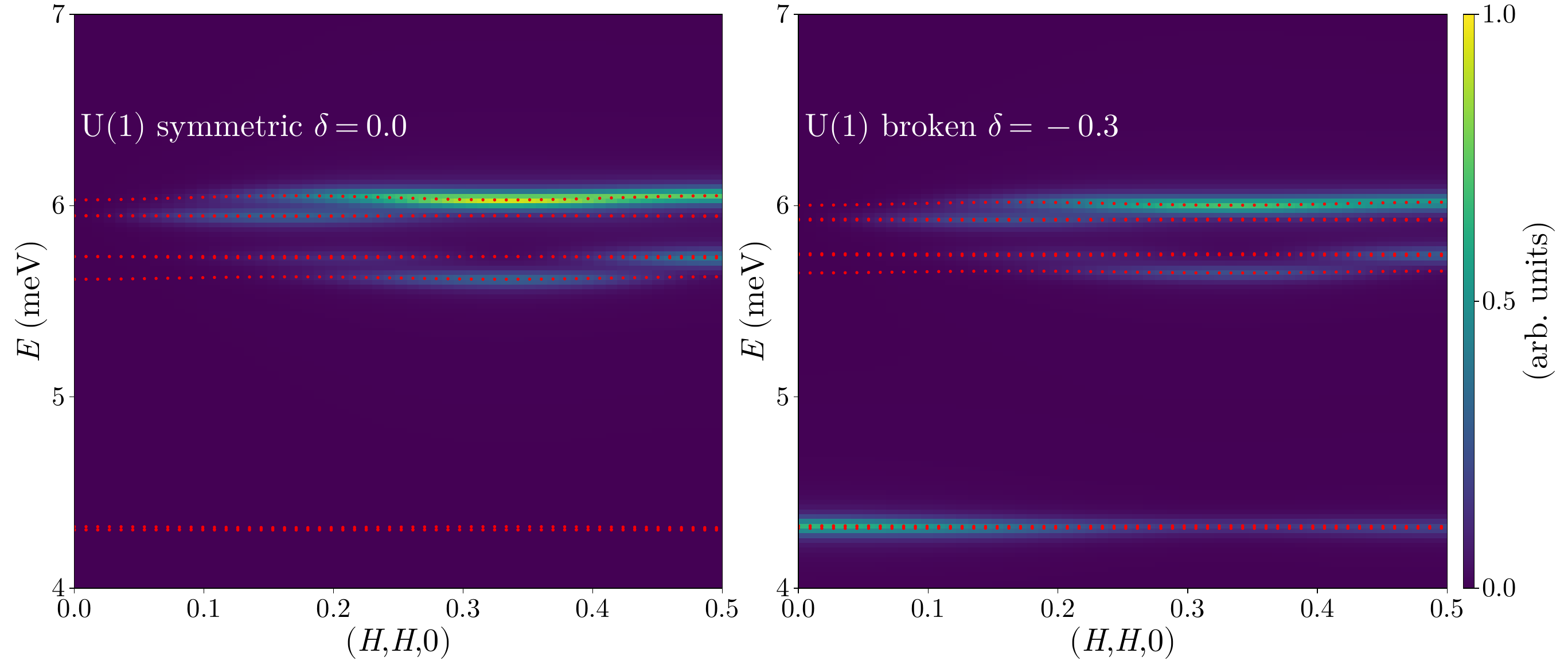}
    \caption{Effect of U(1) symmetry breaking on the INS spectrum of $\kcs$. The red dots indicate the eigenstates of the truncated Hamiltonian, overlaid on the INS intensity calculated using the THED method. The left panel shows results for the U(1)-symmetric case ($\delta = 0$), while the right panel shows the spectrum after introducing a finite U(1)-breaking anisotropy ($\delta \neq 0$). Upon breaking U(1) symmetry, the quadrupolar bound state--previously invisible--gains dipolar character and becomes detectable via INS.}
    \label{fig:U1b}
\end{figure}
As discussed in the main text, the two-magnon bound states shown in Fig.~3{\bf d} are invisible to INS experiments due to their quadrupolar nature. This conclusion, however, relies on the assumption that the model Hamiltonian preserves the U(1) symmetry of global spin rotations along the magnetic field axis, as is the case for Eq.~1 in the main text. We now explicitly break the U(1) symmetry by introducing an additional anisotropy (allowed by symmetry) in the $S^yS^y$ term: 
\begin{equation}
    \mathcal{H} = J \sum_{\langle i,j \rangle} \left(S_i^xS_j^x+(1+\delta)S_i^yS_j^y + \Delta S_i^zS_j^z \right) - g_{cc}\mu_B\sum_i S_i^z.
\end{equation}
The parameter $\delta$ quantifies the strength of U(1) symmetry breaking. When $\delta \neq 0$, $S^z$ is no longer conserved, enabling quadrupolar excitations with $\Delta S^z = -2$ to hybridize with dipolar modes. As a consequence, previously ``dark'' quadrupolar bound states acquire finite dipolar character and become visible in INS measurements.

To illustrate this idea, we adopt the model parameters relevant to $\kcs$ and compute the corresponding INS intensity using the THED method, both with and without the U(1)-breaking term. Figure~\ref{fig:U1b} compares the two cases: the left panel shows the spectrum for the U(1)-symmetric case ($\delta = 0$), while the right panel includes a finite $\delta$ that explicitly breaks the symmetry. As anticipated, breaking U(1) symmetry allows the quadrupolar bound state to acquire dipolar character, leading to finite spectral weight and making it detectable in INS.

Crucially, no corresponding spectral feature is observed at the predicted energy in the experimental data for $\kcs$, strongly suggesting that the system preserves U(1) symmetry to a very good approximation. This observation supports the conclusion that $\kcs$ is accurately described by the Hamiltonian proposed in Refs.~\cite{ZhuM2024a,ZhuM2024b}, even though a U(1)-breaking term is symmetry-allowed. This analysis, made possible by the interpretability and efficiency of the THED method, highlights its power in constraining microscopic Hamiltonians directly from spectroscopic data.

\bibliographystyle{apsrev4-2}
%